\documentclass[twocolumn]{aastex62}

\usepackage{hyperref}
\usepackage{amsmath}

\received{January 1, 2018}
\revised{January 7, 2018}
\accepted{\today}
\submitjournal{ApJ}

\shorttitle{Escape Fraction in Emission Line Galaxies}
\shortauthors{Griffiths et al.}

\def\casgm20{CAS-G-M$_{20}\,$}
\def\m20{M$_{20}\,$}

\begin{document}

\title{Emission Line Galaxies in the SHARDS Hubble Frontier Fields II: Limits on Lyman-Continuum Escape Fractions of Lensed Emission Line Galaxies at Redshifts $2 < z < 3.5$}


\correspondingauthor{Christopher J. Conselice}
\email{conselice@manchester.ac.uk}

\author{Alex~Griffiths}
\affiliation{School of Physics and Astronomy, The University of Nottingham, University Park, Nottingham NG7 2RD, UK}
\author[0000-0003-1949-7638]{Christopher J. Conselice}
\affiliation{Jodrell Bank Centre for Astrophysics, University of Manchester, Oxford Road, Manchester UK}
\author{Leonardo Ferreira}
\affiliation{School of Physics and Astronomy, The University of Nottingham, University Park, Nottingham NG7 2RD, UK}
\author{Daniel Ceverino}
\affiliation{Departamento de Fisica Teorica, Modulo 8, Facultad de Ciencias, Universidad Autonoma de Madrid, 28049 Madrid, Spain}
\affiliation{CIAFF, Facultad de Ciencias, Universidad Autonoma de Madrid, 28049 Madrid, Spain}
\author{Pablo G. P\'{e}rez-Gonz\'{a}lez}
\affiliation{Centro de Astrobiolog\'{\i}a (CAB, CSIC-INTA), Carretera deAjalvir km 4, E-28850 Torrej\'on de Ardoz, Madrid, Spain}
\author{Olga Vega}
\affiliation{Instituto Nacional de Astrof\'{i}sica, \'{O}ptica y Electr\'{o}nica, AP 51 y 216, 72000, Puebla, M\'{e}xico}
\author{Daniel Rosa-Gonz\'{a}lez}
\affiliation{Instituto Nacional de Astrof\'{i}sica, \'{O}ptica y Electr\'{o}nica, AP 51 y 216, 72000, Puebla, M\'{e}xico}
\author{Anton M. Koekemoer}
\affiliation{Space Telescope Science Institute, Baltimore, MD 21218, USA}
\author{Danilo Marchesini}
\affiliation{Physics and Astronomy Department, Tufts University, 574 Boston Ave, Medford, 02155 MA, USA}
\author{Jos\'{e} Miguel Rodr\'{i}guez Espinosa}
\affiliation{Departamento de Astrof\'{i}sica, Universidad de La Laguna, E-38206 La Laguna, Spain}
\author{Luc\'ia Rodr\'iguez-Mu\~{n}oz}
\affiliation{Dipartimento di Fisica e Astronomia, Universit{\`a} di Padova, vicolo dell'Osservatorio 3, I-35122 Padova, Italy}
\author{Bel\'{e}n Alcalde Pampliega}
\affiliation{European Southern Observatory (ESO), Alonso de C\'{o}rdova 3107, Vitacura, Casilla 19001, Santiago de Chile, Chile}
\affiliation{Departamento de F\'{i}sica de la Tierra y Astrof\'{i}sica, Faultad de CC F\'{o}sicas, Universidad Complutense de Madrid E-2840 Madrid, Spain}
\author{Elena Terlevich}
\affiliation{Instituto Nacional de Astrof\'{i}sica, \'{O}ptica y Electr\'{o}nica, AP 51 y 216, 72000, Puebla, M\'{e}xico}

\begin{abstract}
We present an investigation on escape fractions of UV photons from a unique sample of lensed low-mass emission line selected galaxies at $z<3.5$ found in the SHARDS Hubble Frontier Fields medium-band survey.  We have used this deep imaging survey to locate 42 relatively low-mass galaxies, down to log (M$_{*}/M_{\odot}) = 7$, between redshifts $2.4 < z < 3.5$ which are candidate line emitters. Using deep multi-band Hubble UVIS imaging we investigate the flux of escaping ionizing photons from these systems, obtaining $1\sigma$ upper limits of {$f^{rel}_{esc}$} $\sim 7$\% for individual galaxies, and $<2$\% for stacked data. We measure potential escaping Lyman-continuum flux for two low-mass line emitters with values  at $f^{\rm rel}_{\rm esc} = 0.032^{+0.081}_{-0.009}$ and  $f^{\rm rel}_{\rm esc} = 0.021^{+0.101}_{-0.006}$, both detected at the $\sim3.2\sigma$ level.  A detailed analysis of possible contamination reveals a $<0.1$\% probability that these detections result from line-of-sight contamination.   The relatively low Lyman-continuum escape fraction limit, and the low fraction of systems detected, is an indication that low-mass line emitting galaxies may not be as important a source of reionization as hoped if these are analogs of reionization sources.  We also investigate the structures of our galaxy sample, finding no evidence for a correlation of escape fraction with asymmetric structure.

\end{abstract}



\section{Introduction}

The universe started off as a hot plasma consisting mostly of free electrons and protons.   Roughly 370,000 years after the big bang, the universe had cooled enough such that these electrons and protons combined together to form neutral hydrogen atomic gas opaque to electromagnetic radiation. Today, we know that the universe is once again, for the most part, ionized through a process called 'reionization'.  The large number of photons which have been emitted by galaxies through AGN and star-formation have produced a universe that is now transparent, and in which the average hydrogen atom is ionized. The era in which the bulk of this transition occurred is known as the epoch of reionization (EoR). Its history, and the physical processes responsible for it are a major field of astronomical investigation \citep[e.g.][]{Ouchi2010,Robertson2013,Duncan2015,Smith2020}.

We have some insights into how this process of transition developed based on both the spectra of distant quasars at $z > 6$, and the properties of the cosmic microwave background. Quasar spectra, due to the Gunn-Peterson effect, show significant absorption in light emitted below the Lyman-limit, the energy limit in which photons are able to ionize hydrogen \citep[e.g.][]{Fan2006, flury2022}.  This puts a lower limit on reionization as having occurred at $z > 6$. The upper limit of this epoch is determined by the Thompson scattering of electrons from cosmic microwave background observations. 

The latest Planck results \citep{Planck2018} suggest that the reionization of the Universe happened relatively fast and late, with a mid-point redshift of $z = 7.7 \pm 0.7$. These results are consistent with theories whereby reionization is driven primarily by massive stars, and possibly active galactic nuclei (AGN) in low-mass galaxies \citep[e.g.][]{Duncan2015,Robertson2015,Parsa2018, naidu2020}.  However, we lack precise, direct measurements of the amount of ionizing photons escaping from galaxies of different masses, especially for those at lower masses which are the most common.

These low-mass galaxies within the EoR are difficult to study due to their faintness, and the opacity of the intergalactic medium (IGM) at high redshifts \citep[e.g.][]{Vanzella2018}. Searches for these systems are often aided by the strong gravitational lensing powers of galaxy clusters \citep[e.g.][]{Bhatawdekar2019} and now JWST \citep[e.g.,][]{Adams2022, Trussler2022, Bradley2022}.   High redshift samples are still however often biased towards the most luminous sources whose light is able to permeate through the high column density of the IGM.  This includes bright quasars and galaxies with strong Lyman-alpha emission. It therefore remains possible that the bulk of galaxies responsible for reionization have yet to be discovered  or properly identified. 
  
Recent works have suggested that star forming galaxies alone may not be sufficient to drive reionization. With new optical depth estimates from Planck \citep{Planck2018}, the need for a large ionizing background at high redshifts is reduced. Combined with the discovery of early ($z > 4$) populations of faint AGN \citep{Glikman2011,Giallongo2015}, this has prompted a reassessment of the contribution of AGN to cosmic reionization. Observational evidence \citep{Giallongo2015,Madau2015} and theoretical studies \citep{Yoshiura2017,Torres2020} show that AGN are possibly able to provide a significant source of ionizing flux, while investigations of known Lyman-continuum (LyC) leakers indicate potential AGN components \citep[e.g.][]{Jia2011,Prestwich2015,Kaaret2017,Grazian2018}.  However, other studies such as e.g., \citet{Parsa2018} show that AGN may in fact not be needed, nor an important aspect in reionization.   There is therefore significant debate in this matter that further observations can help address.

As well as the internal feedback from AGN and star formation, external influences, such as the merger history can play a crucial role in the production of escaping ionizing photons. This is particularly important in low-mass halos \citep[e.g.][]{Chen2014} in which a merger event can readily expel large fractions of gas from the ISM, resulting in lower column densities in which ionizing radiation can propagate more freely. For more massive galaxies, it has even been shown that fast accretion shocks associated with gravitational infall of baryons due to events such as mergers can also contribute to reionization \citep[e.g.][]{Dopita2011,Wyithe2011}.

In the last few decades, significant effort has gone into searching for sources of escaping Lyman-continuum radiation over various redshifts, however many studies report non-detections, yielding mostly upper-limits for individual sources. At low redshifts (0 $<$ z $<$ 3) \citet{Leitherer2016} and \citet{Puschnig2017} report detections in local starbursts, while a number of compact star forming galaxies selected by their high nebular oxygen ratios have been shown to exhibit non-negligible escape fractions \citep[e.g.][]{Izotov2016, Verhamme2017}. At intermediate redshifts, \citet{Bian2017} detect Lyman-continuum emission from a gravitationally lensed compact dwarf galaxy at  $z \approx$ 2.5.  It is still however largely the case that galaxies with unambiguous escaping Lyman-continuum emission are relatively rare.

Furthermore, this is a difficult measurement to make as at higher redshifts light from foreground galaxies along the line of sight can be a  significant   source of contamination \citep[e.g.][]{Bridge2010,Vanzella2012,Siana2015, mostardi2015}.  A high IGM opacity also makes individual measurements challenging, which increases as we probe higher redshifts. However, it is possible to place constraints on reionization through the study of galaxy populations while making assumptions about the IGM clumping factor. \citet{Finkelstein2012} put limits on the escape fraction by investigating the luminosity density of a sample of 483  galaxies at $z = 6-8$ selected from the CANDELS fields \citep{Grogin2011, Koekemoer2011}. Other studies, such as \citet{Grazian2012} use CANDELS data to examine the size-luminosity relation of 153 Lyman break galaxies at $z \sim 7$ to infer limits on their contribution to reionization. These studies typically find that the luminosity function of star forming galaxies must extend down to M$_{UV} \sim -13$ in order to fully reionize the universe by $z=6$.
 
To circumvent some of these issues, in this paper we examine lower redshift analogues of the galaxies that might play a key role in the reionization of the universe - those of low mass or dwarf systems. However, this does not come without its challenges. These dwarf galaxies, while being the most abundant type of galaxy in the universe, are difficult to detect due to their low-mass and luminosity, which both contribute to their faintness and low surface brightness. Developments in data analysis techniques and deep survey strategies have helped to provide a further insight into this population via group/cluster \citep[e.g.][]{Ferrarese2012,Kondapally2018} and field \citep[e.g.][]{Leisman2017} studies.

Galaxies at redshifts $z \approx 3 - 4$ are within the ideal redshift range for the identification of LyC emission \citep[e.g.][]{Steidel2018}. At z $\gtrsim$ 3, the Lyman-limit is redshifted into the rest-frame U-band, where the sky  background is dark and the atmosphere is transparent, allowing for observations from ground-based telescopes. These sources also benefit from low IGM opacities. However, a limited number of LyC emitters have been identified at these redshifts \citep[e.g.,][]{deBarros2016,Shapley2016,Vanzella2016,Steidel2018,marchi2018, fletcher2019, naidu2020,nakajima2020, pahl2021} as searches are complicated by low redshift interlopers which can result in false detections, requiring high resolution imaging and spectroscopy for confirmation.   To understand which galaxies might be analogs of the sources producing reionization we explore different types of galaxies, including lower mass ones such as those discussed in this paper.

In order to establish how important these lower mass sources are as a driver for reionization, it is necessary to determine the net fraction of ionizing radiation that is able to escape the high column density {H}{I} gas surrounding these galaxies. This can be achieved by measuring the escape fraction, $f_{esc}$ \citep{Steidel2001}, a parameterization of the fraction of Lyman-continuum ionizing photons which are not absorbed by either the interstellar medium (ISM) surrounding young stars, the host galaxy's circumgalactic medium, or the intergalactic medium. 

As such, in this paper we measure the escape fractions in a sample of lensed emission line galaxies identified via medium-band observations, utilising the strong lensing features of two Hubble Frontier Fields clusters. Our data is unique in that our sources are discovered in very deep medium-band imaging from the SHARDS-FF survey obtained with the Gran Telescopio de Canarias (GTC), with follow up UV imaging from the Hubble Space Telescope. 

This paper is outlined as follows.  In Section~\ref{sec:obs} we provide an overview of the observations and ancillary data, as well as details of the medium-band selection procedure. We calculate escape fraction upper limits for stacked, and individual sources, and detail our structural classification in Section~\ref{sec:results}. Finally, we summarise our main results and draw conclusions in Section~\ref{sec:discussion} and \ref{sec:conclusion}. Throughout this paper we adopt a $\Lambda$ cold dark matter cosmological model with $\Omega_{\Lambda} = 0.7$, $\Omega_{\textrm{M}} = 0.3$ and $H_0 = 70$ km s$^{-1}$ Mpc$^{-1}$. All magnitudes are given in the AB system \citep{Oke1974}.

\section{Observations and Data Reduction}
\label{sec:obs}

The present analysis is based on new medium-band observations of two of the Hubble Frontier Fields (HFF) galaxy clusters \citep{Lotz2017}, Abell 370 and MACS J1149.5+2223 and their corresponding parallel fields. We couple these medium-band observations with the multi-wavelength photometric catalogues made available through the HFF-DeepSpace project \citep{Shipley2018}.   We focus on candidate line emitters as a subsample of star forming galaxies, and those that are likely in their early stages.  It is possible that these systems are analogs of the sources that reionize the universe given their low mass and line emitting nature.  We describe these samples below in some detail.

\subsection{HST and Medium-band imaging}

\begin{figure*}
	\includegraphics{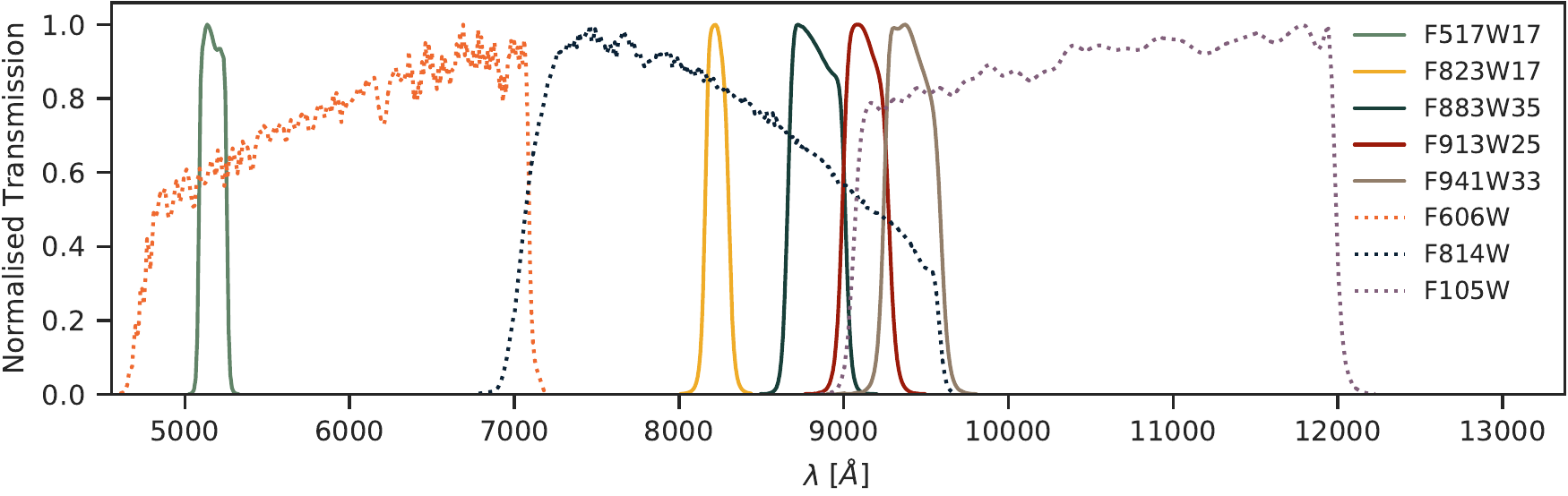}
	\caption{Transmission curves of the medium-band SHARDS filters (F517W17, F823W17, F883W35, F913W25 and F941W33) and broad-band HST filters (F606W, F814W, F105W) used for the selection of emission line galaxies. All curves have been individually normalised by their maximum transmission.  We use the SHARDS medium band filters to select our line emitter candidates in comparison to the broad-band HST filters (F606W, F814W and F105W), whose filter transmission curves are also shown.}
	\label{fig:filters}
\end{figure*}

Medium-band imaging data of the Abell 370 and MACS J1149.5+2223 clusters (hereafter M1149 and A0370) are obtained from the ongoing SHARDS Frontier Fields survey (SHARDS-FF; PI:P\'erez-Gonz\'alez). Observations simultaneously cover the cluster and their respective parallel fields in a single pointing. At the time of this study, observations of A0370 have been carried out with four SHARDS filters (F517W17, F823W17, F913W25 and F941W33), while M1149 has been observed in three filters (F883W35, F913W25 and F941W33).   The SHARDS-FF observations are performed with the Optical System for Imaging and low-Intermediate-Resolution Integrated Spectroscopy (OSIRIS) instrument, at the 10.4 m Gran Telescopio de Canarias at the Observatorio del Roque de los Muchachos, in La Palma. A total of 240 hours of time has been granted to the SHARDS-FF program with observations beginning in December 2015.   The wavelength response of these filters, in comparison to the HST comparision filters is shown in Figure~\ref{fig:filters}.

Individual images are reduced using a dedicated OSIRIS pipeline \citep{Perez2013}. The pipeline performs bias subtraction and flat fielding as well as illumination correction, background gradient subtraction and fringing removal. Additionally the pipeline implements World Coordinate System (WCS) alignment which includes field distortions, two-dimensional calibration of the pass-band and zero point and stacking of individual frames. 

In order to select emission line galaxy candidates, we require well calibrated multi-wavelength broad-band catalogues. For this reason we utilise the data made available as part of the HFF-DeepSpace project \citep{Shipley2018}. These data combine up to 17 ACS/WFC3 filters with ultra-deep Ks band imaging and Spitzer-IRAC, when available. We also include UVIS HST observations wthin the filters F225W, F275W and F336W for the M1149 cluster, and F275W and F336W for the A0370 cluster. We also utilize the F160W observations within WFC3 for morphological measurements. 

Catalogues we use also include photometric redshifts, lensing magnification factors as well as original imaging data, models, and calibration information, providing an ideal ancillary dataset for candidate selection. We note here that HFF-DeepSpace observations cover only a fraction of the area surveyed by SHARDS-FF. Considering the cluster and parallel field, this roughly constitutes $\sim$35\%~and $\sim$38\% of the total SHARDS-FF coverage for A0370 and M1149 respectively. 

\subsection{Lensing Magnification}

In order to provide accurate measurements of stellar mass and luminosity derived properties, the strong gravitational lensing effects induced by these massive galaxy clusters need to be accounted for. For this, reliable lensing models are required. Fortunately, for each of the six HFF clusters seven independent teams have undertaken  work to produce lens models, each employing various methods and initial assumptions. The magnification factors derived from these models are provided in the HFF-DeepSpace dataset, and full details of their derivation can be found in Section 5.5 of \citet{Shipley2018}. For further information on the lensing models, we refer the reader to the MAST website\footnote{https://archive.stsci.edu/prepds/frontier/lensmodels/}. To demagnify stellar masses and exclude any outliers, we take the median magnification factor ($\mu$, provided in later Tables) from all of the lens models available in the HFF-DeepSpace data. 

We find an average lensing magnification for the sample of $<2.98>$ across all galaxies in our sample.   The distribution is however very skewed with the mode $= 1$, where many galaxies are not demagnified at all, and the maximum value is 19.48.  The mean absolute deviation (MAD) of the magnification values is 2.22.    We use the individual magnification factors to demagnify the fluxes for our sample from which we calculate the stellar mass and UV magnitudes for our systemes which we use later in this paper.  All of these values, which we use throughout, have been demagnified by each galaxy's magnification factor, unless otherwise stated. 

\subsection{Selection of Line-Emitters}
\label{sec:selection}

We focus this paper on candidate line emitters at $z > 2$, as we can with some confidence conclude they are at these redshifts.  However, we will need spectroscopy to confirm that these single lines are within our redshift of interest.  We examine these systems in detail as line emitters might also be more likely to contain a high escape fraction and/or higher levels of ionizing photons \citep[][]{stark2015, Naidu2022, Schaerer2022}, as opposed to normal star forming galaxies without line emission. Part of the reason is that it is not yet clear whether or not galaxies which are line emitters are always more likely to produce Lyman-continuum emission that escapes.  For example, \citet{Wang2021} find fewer Lyman-continuum leakers as a function of measured [SII] line equivalent width, implying that galaxies with escaping LyC flux are more likely found at lower line fluxes. This is thus an important question that deserves more attention.

To select for emission line galaxy candidates, the medium-band data is flux calibrated to match the existing broad-band catalogues. In this section we provide a brief summary of the data calibration and candidate selection relevant for this paper.  A full discussion and details for the entire SHARDS Frontier Fields data is provided in Griffiths et al. (2021).   

Firstly, we calculate aperture photometry of all individual medium-band mosaics using SExtractor \citep{Bertin1996} in dual image mode, using deep detection images from the HFF-DeepSpace dataset. We do this carefully by matching PSFs and aligning the data between the bands. First we match all observations to the PSF of the SHARDS F883W filter, which has a seeing of $\sim 1.0$\arcsec. We do this by deriving convolution kernels for
each band using the PSFs retrieved from the SHARDS
and Hubble data sets (see \citet{Griffiths2021} for more detail). Once this is done, we then use 2 arcsec apertures to recalculate photometry in SExtractor’s
dual image mode, utilizing the deep detection images from the HFFDeepSpace data (see Shipley et al. 2018, section 3.3).   We  correct for any flux falling outside of the 2 \arcsec apertures, by deriving total flux values. Aperture photometry is adjusted by
applying a correction factor calculated for each galaxy, although in practice this is small.

After converting aperture fluxes to total flux, we correct for galactic extinction and standardise all zero-points to 25, as our photometry is adjusted for empirically determined zero-point errors \citep[see][for further details]{Skelton2014}.

To perform candidate selection we first calculate photometric redshifts using our new medium-band data with the HST data using EAZY\footnote{\url{https://github.com/gbrammer/eazy-photoz/}} \citep{Brammer2008}. We run EAZY on an object-to-object basis due to the spatially varying central wavelengths of the SHARDS-FF filters. The geometric effects of the SHARDS-FF observations are due to the incidence angles of the GTC/OSIRIS light beam, which causes the effective central wavelength of the filter to vary as a function of location within the image. The central wavelength is calculated for each object in order to provide more accurate photometric redshift estimates. We further use the stellar population synthesis code, FAST\footnote{\url{http://w.astro.berkeley.edu/~mariska/FAST.html}} \citep{Kriek2009} to estimate stellar masses, star formation rates, ages and dust extinctions, based on the photometric redshifts from the EAZY output. 

\begin{figure*}
	\includegraphics[width=\textwidth]{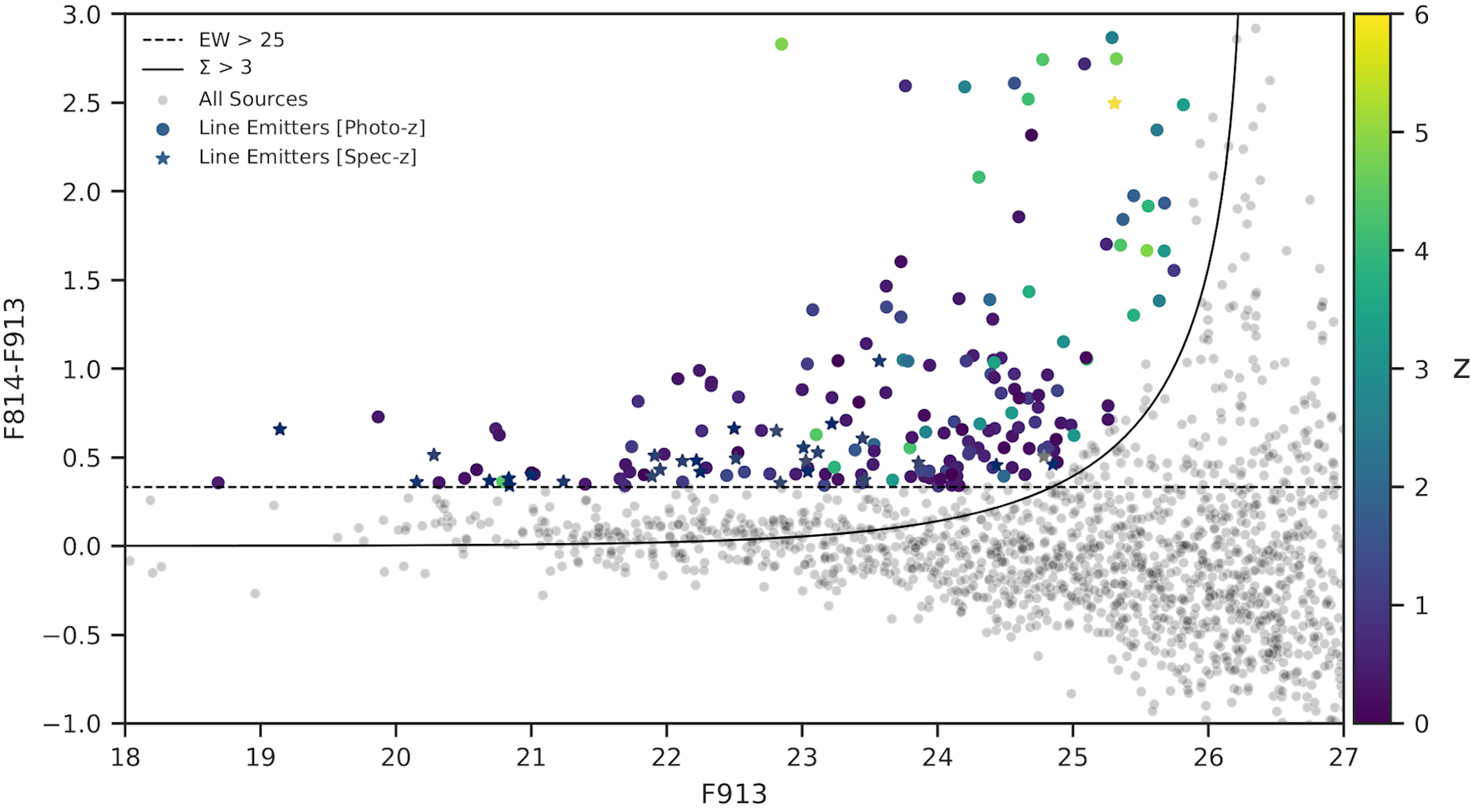}
	\caption{An example of the selection method for emission line candidates from the A0370 cluster and parallel field through the F913W medium-band filter. This process is repeated with all filter and field combinations in order to identify emission line candidates over all redshifts, for more detail see \citet{Griffiths2021}. The sub-sample used in this paper is selected based on the HST/UVIS filter and the redshift combination detailed in Table~\ref{tab:candidates}. Small grey points show all sources in the fields, while larger points are coloured by redshift and represent candidates in which the selection criterion is met.  Stars represent sources in which spectroscopic redshifts are available.  Symbols that are in gray do not satisfy our selection criteria. The solid line represents the average $\Sigma$ = 3 colour significance, and the dotted horizontal line shows observed the observed EW = 25 \AA.}
	\label{fig:select}
\end{figure*}

For the  selection of emission line candidates, amongst galaxies in our redshift range of interest, we employ a two parameter selection method which has been well established in previous studies \citep[e.g.][]{Matthee2015,Santos2016,Sobral2017}. These criteria assures that the objects selected show real colour excesses due to an emission line, and not from random scatter or measurement uncertainties. The first of these criteria is to set a lower limit on the observed emission line equivalent width (EW). We follow the methods used in previous searches and set an observed EW cut of 25\AA~\citep[e.g.][]{Ouchi2008} to remove sources with little or no excess in the medium-band. The equivalent width is calculated via the equation: 

\begin{equation}
	EW = \Delta\lambda_{MB}\frac{f_{MB}-f_{BB}}{f_{BB}-f_{MB}\frac{\Delta\lambda_{MB}}{\Delta\lambda_{BB}}}
    \label{eq:ew}
\end{equation}

\noindent where MB stands for medium-band, BB stands for broad-band, such that $\Delta\lambda_{MB}$ and $\Delta\lambda_{BB}$ are the widths of the medium and broad-band filters respectively and $f_{i}$ are the flux densities. The second criteria we use for the line emitter selection is the excess significance \citep[$\Sigma$, e.g.][]{Bunker1995}.  This is used to quantify if the flux excess is real, or due to random scatter. The excess significance is calculated as \citep{Sobral2013}: 

\begin{equation}
	\Sigma=\frac{1-10^{-0.4(BB-MB)}}{10^{-0.4(ZP-MB)}\sqrt{\pi r_{ap}^2(\sigma_{px,BB}^2-\sigma_{px,MB}^2))}}
    \label{eq:sig}
\end{equation}

\noindent where $MB$ and $BB$ are the medium and broad-band magnitudes respectively, $ZP$ represents the normalised zero-point (25) and $r_{ap}$ is the aperture radius in pixels. The root-mean-square (rms) of the background pixel values, $\sigma_{px}$, are estimated by randomly placing empty apertures across the respective images. A selection criterion of $\Sigma > 3$ is used to classify sources as potential line emitters \citep{Sobral2013}. We show the visual representation of this selection process for the A0370 cluster and the F913W medium-band filter in Figure~\ref{fig:select}.

\begin{table}
\caption{Listed here is the filter of observation, the number counts, and redshift range for the emission line galaxies with Lyman-continuum data we examine. The last two columns lists the number of sources identified within each field and band.}

\label{tab:candidates}
\centering
\begin{tabular}{ccccc}
  \hline
  ionizing  &  Non-ionizing  &  Redshift Range  & \multicolumn{2}{c}{\# Candidates}  \\
  (LyC) & (UV) & z & M1149 & A0370             \\
  \hline
  F275W  &  F606W  &  2.4 - 2.9  &  4  &  18   \\
  F336W  &  F606W  &  3.0 - 3.5  &  3  &  17   \\
  \hline
 \end{tabular}
\end{table}

To select potential Lyman-continuum emitters from our overall sample of emission line galaxies, we utilise HST/UVIS observations: this includes the filters: F225W, F275W, and F336W for the M1149 cluster, but is limited to only F275W and F336W for the A0370 cluster (note that these data are not available for the parallel fields). From the total sample of emission line candidates, these filters permit a search for potential Lyman-continuum emitters over the redshift range z~$\approx 2 - 3.5$, yielding a total sub-sample of  62 candidate objects that were originally in our first cut sample.  Ultimately we narrow this down to 42 and in Table~\ref{tab:candidates} we provide a full breakdown of the filter dependent redshift ranges and final emission line number counts.  

The redshifts of our final objects were selected so as to include LyC radiation within each filter of observations.  We also are careful to remove objects with redshifts that might be affected by red leaks.   We select these redshift ranges to ensure as much as possible that non-ionizing continuum radiation at $> 912$\AA is $< 1$\% of the light we detect in the UVIS filters. From our calculations, which mirror that of \citet{Smith2020} we determine that at most only a single object we observe will have more flux contamination than this due to the red leak.  Neither of our individual detections, however, would be affected by this.

\begin{figure}
	\includegraphics[width=\columnwidth]{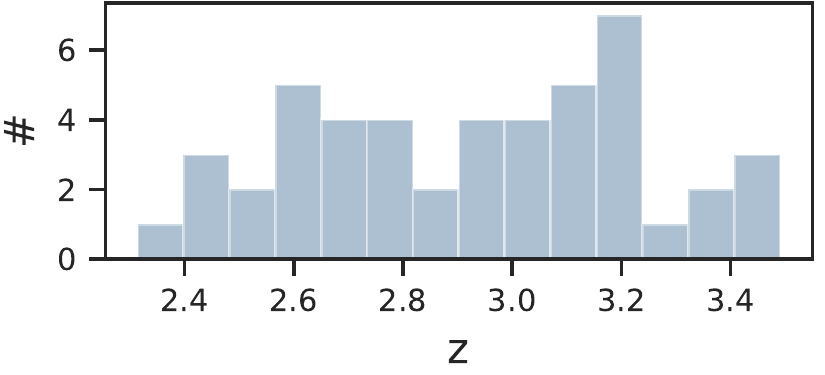}
	\caption{Photometric redshift distribution of line emitter candidates from our sample of photometrically selected sources at $2 < z < 3.5$. }Photometric redshifts are obtained with EAZY with the inclusion of all SHARDS medium-band photometry with HFF-DeepSpace catalogues.
	\label{fig:zdist}
\end{figure}

\begin{figure}
	\includegraphics[width=\columnwidth]{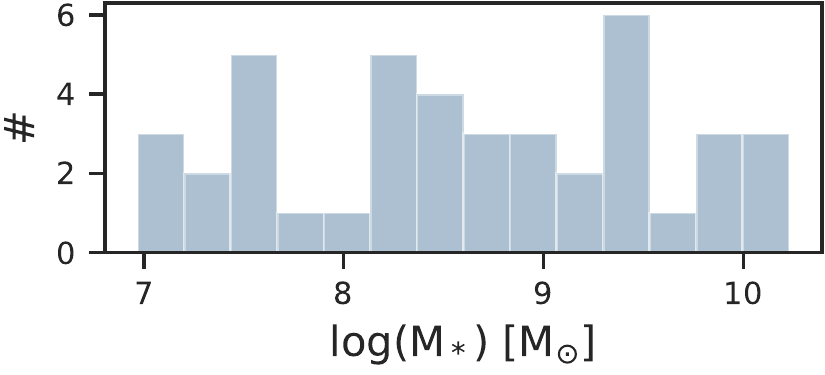}
	\caption{Demagnified stellar mass distribution of the line emitter candidates from our sample. Stellar masses have been corrected for lensing magnification and are obtained using our updated photometric redshifts and FAST SPS modelling, implementing all SHARDS medium-band photometry.}
	\label{fig:z_vs_mass}
\end{figure}

The stellar mass and redshift distributions of this final sub-sample we use in our detailed analyses for LyC escaping light are shown in Figures~\ref{fig:z_vs_mass} and~\ref{fig:zdist} respectively.  As mentioned early, these values have been demagnified as outlined in \S 2.2.

\subsection{Spectroscopic redshifts and selected emission lines}

When available, we utilise spectroscopic redshifts obtained from the HFF-DeepSpace catalogues, as well as from VLT/MUSE observations via the work of \citet{Lagattuta2019}. Spectroscopic redshifts in the HFF-DeepSpace catalogues are compiled from the literature (see Section~5.1 of \citet{Shipley2018}, while \citet{Lagattuta2019} investigate the MUSE observations obtained as part of programme 096.A-0710(A) (PI: Bauer) with a combination of targeted extractions based on HST imaging, and a `blind' emission line search. Spectroscopic redshifts are available for nine of our candidates, comparing these values to our photometric redshift estimates we find our photo-zs are robust with an average $\Delta z/(1+z)= 0.04$, with the exception of object ID:2330 which is found to be contaminated by a nearby bright source ($z_{phot} = 2.06,~z_{spec} = 3.81$).  This is a higher redshift accuracy that what was found in the overview paper of \citet{Griffiths2021}.  This is due to the fact that we are studying a sample of galaxies at a well defined redshift range, whereby the phtometric redshifts are more straightforward to measure than at lower or higher redshifts. Given that \citet{Griffiths2021} spans the entire redshift range of possible line emitters, it is natural for the accuracy of photometric redshifts to be lower for this larger range.

The combination of available HST/UVIS data and our medium-band selection method provides the unique opportunity to probe a rarely studied population of lower mass galaxies between $2 < z < 3.5$ which have emission lines. With the exception of a few systems, such as galaxy ID:1629, which is selected by the Ly$\alpha$ 1216\AA~line, our sample is identified via lines typically present in the galaxies thought to be analogues of those which contribute significantly to reionization at $z > 6$. These lines possibly include rest-frame UV emission lines such as CIV, {N}{V} 1240\AA, {C}{III}] 1909\AA, {Mg}{II} 2798\AA, various Fe lines, and the 2175 \AA\ bump.  These lines are  present in AGN, and in low metallicity star forming regions where hard spectra from low-metallicity stars can excite the higher ionization lines.  {\ However, without spectroscopy we cannot confirm the exact line identification for our candidates.  Our sample is ultimately a photometric redshift selection of galaxies detected with line emission with a redshift that places them within the range where they can have their Lyman-continuum light examined between $2 < z < 3.5$. 

We remove from our sample any systems where there are no possible corresponding emission lines near the measured photometric redshift. We can better understand our photometric redshift distribution if we consider our photo-z accuracy of  $\delta z / (1+z) \sim 0.04$, resulting in a $\delta z \sim 0.16-0.18$ uncertainty, within our redshift range. This can account for some of the blurred distribution seen in Figure~4, where we would normally expect to find discrete redshifts based on where known lines would be located in our filters.}

Thus, our sample provides a method for identifying candidate faint and low-mass galaxies at these redshifts.   Figures~\ref{fig:zdist} and~\ref{fig:z_vs_mass} show the distribution of stellar masses and photometric redshifts for our line emitter sample that we study throughout this paper.  It is important to note that without further spectroscopic follow up it is difficult to exactly identify the appropriate emission line(s) responsible for the observed flux excess at these redshifts. This is beyond the scope of this paper, and as such we make no inferences about the properties of our sample galaxies based on this selection, although it is possible that a fraction are active AGN.  Follow-up spectroscopy of our sources will be required to confirm their exact redshifts, determine the observed emission line identity, and to confirm any AGN nature.

\section{Results}
\label{sec:results}

For the remainder of this paper we discuss the Lyman-continuum limits for our sample, and examine two tentative detection of Lyman-continuum emitters from our photo-z/line emission sample.   We first give a description of our formalism, and then discuss the individual Lyman-continuum detected objects. We then anaylse the stacked limits for the sample itself based on this formalism.

\subsection{Lyman-continuum Escape Fraction Formalism}
\label{sec:ef}

The UV escape fraction of high redshift galaxies can be derived from differential measurements of the ionizing flux (LyC, $\lambda_{\textrm{rest}} \leq 912$\AA) compared to non-ionizing (UV, $\lambda_{\textrm{rest}} \simeq 1500$\AA) emission.  However, as the intrinsic spectral energy distribution (SED) of galaxies is not typically known, and the LyC from young stars is readily attenuated through the ISM and IGM, relative escape fractions are usually derived from observations while making assumptions on the intrinsic flux ratio based on models. 

The relative escape fraction, $f_{esc}^{rel}$, was first introduced by \citet{Steidel2001} and is defined as the fraction of escaping LyC photons divided by the fraction of escaping non-ionizing (UV) photons.  Following \citet{Shapley2006} and \citet{Siana2007}, the relative escape fraction is derived from the observed flux density ratio $(f_{UV}/f_{LyC})_{obs}$ via the equation:

\begin{equation}
	f_{esc}^{rel}=\frac{\left(f_{UV}/f_{LyC}\right)_{int}}{\left(f_{UV}/f_{LyC}\right)_{obs}}{\rm exp}\left(\tau_{IGM}\right)
    \label{eq:relEF}
\end{equation}

\noindent where $(f_{UV}/f_{LyC})_{int}$ is the intrinsic flux ratio and exp$(\tau_{IGM})$ is the inverse of the redshift dependent IGM attenuation of LyC by neutral {H}{I}. 

\subsection{Individual limits}

Using the methods outlined in Section~\ref{sec:ef} we measure relative escape fractions for our emission line selected galaxies in two of the frontier fields clusters at $2 < z < 3.5$. We successfully measure rest-frame ionizing flux densities of more than 2$\sigma$ significance in just two galaxies from our sample (as we will discuss below), suggesting the detection of leaking Lyman-continuum emission within at least these two systems. We describe our measurement, tests for contamination, and ultimately describe the escape fraction limits below.

\subsubsection{Measurements}

We measure the escape fractions of individual galaxies utilizing calibrated HST images. These images are cleaned of cosmic rays, then background subtracted, PSF matched, and the brightest cluster galaxies  are modelled and subtracted out. We utilise all available HST/UVIS bands (filters F225W, F275W and F336W with $1\sigma$ depths of $\sim$ 29 mag), and select complementary imaging in the F435W and F606W bands in order to sample the non-ionizing emission at a rest-frame UV wavelength of $\lambda \sim 1500$\AA. We define redshift ranges and filter matches such that the Lyman-limit falls outside of the filters bandpass. This is necessary in order to avoid contamination by non-ionizing emission in the LyC measurements. The full details of the filter combinations and redshift ranges see Table~\ref{tab:candidates}.

 To measure the fluxes from our sample we initially take total flux values from the HFF-DeepSpace catalogues for the corresponding bands, in which the well calibrated multi-wavelength photometry provides an estimate for the LyC to UV flux ratios. The HFF-DeepSpace photometry is performed in SExtractor dual image mode using a deep, stacked detection image created from the F814W, F105W, F125W, F140W and F160W bands \citep[see][for more details]{Shipley2018}. However, in the majority of cases we find no detections of LyC flux within the catalogue. Instead, we utilise the segmentation maps from these deep detection images to mask objects and obtain reliable background estimates, from which 1$\sigma$ upper flux limits are estimated.    

We do this as the LyC photons may be emitted from sub-regions of a galaxy.  That is, the LyC flux may not be cospatial with the non-ionizing emission or rest-frame UV flux within a galaxy \citep[e.g.,][]{Nestor2011}. Because of this, the LyC emission may not always be associated with non-ionizing emission, as described in \citet{Vanzella2012}. Thus, in order to obtain accurate flux ratio estimates, we conduct careful photometric measurements within the same spatial region of the ionizing and non-ionizing images. As all images are matched in pixel scale and PSF, we map isophotes from the deep detection images to the measurement images in order to obtain spatially correlated isophotal fluxes over both the LyC and UV bands.

\subsubsection{Contamination Tests}

Contamination of foreground galaxies near candidate UV continuum emitters is an important aspect that must be considered, as this can easily mimic actual Lyman-continuum detections \citep[e.g.,][]{Nestor2011, Vanzella2012, mostardi2015}. The reason is, if there are other galaxies apparently close in projection, but at different redshifts, we could easily mistake the light from those systems as part of the object itself.  

We deal with this issue in a few ways, which we explain in detail in this section.   We first analyse by eye and investigate all galaxy images closely in order to check for sample contamination.   All of our final Lyman-continuum detected systems are isolated with a well defined structure, and thus do not show any evidence for peculiar structures that could be due to intervening galaxies.  In fact, we find, and remove from our sample, a total of 20 false-positive interlopers, including sources which fall outside of the HST/UVIS pointings and those with a bright object nearby, which we do not further consider.  

 We also investigate the likelihood of contamination in the Lyman-continuum filters by investigating the number density of faint sources detected in the UV filters across our field that, in principle, could result in contamination from an overlap with our sources.  To do this, we calculate the average magnitude in both UV bands for our sample of line emitters from which we investigate the LyC properties, both through the direct detections and the non-detections which we stack (Table~2).  
 
 We find that the average magnitude for detected systems in the F275W and F336W bands is $\sim 27.5$. We make the very reasonable assumption that a contaminating galaxy would have to be at about this magnitude, or fainter, to be considered as an interloper or contamination, otherwise we would be able to visually remove these systems.  In the F275W band we find that the number density of these sources, and fainter, is 14-20 galaxies per arcmin$^{-2}$, with a similar number density in the F336W band.  This results in effectively $\sim$ 14.1$\pm3.6$ sources per square arcmin.  Our two galaxies have LyC fluxes isolated to within around 0.2\arcsec radius aperture in size. Using this fact we calculate the likelihood of a chance superposition based on the aperture in which these detections are made.
 
 There are on average four of these 0.2\arcsec~radius regions within the larger search for LyC emission centered on each galaxy.  We thus calculate the likelihood that the detected flux arises from an alternative non-related source, given the number density of background galaxies calculated in the previous paragraph.  Based on this, we calculate a probability of 1 in 1400 ($< 0.1$\%) that a single LyC detection is a chance superposition with a foreground or background source.   This is certainly smaller than our total number of sources, and is not high enough to account for our two detections.   Thus, we conclude that due to the small physical sizes of our system, and the relative paucity of sources in the field at these magnitudes, that chance superposition is unlikely to be the cause of our two candidate detected LyC fluxes.

\subsubsection{Escape Fraction Calculation and IGM Absorption}

Ideally, intrinsic flux ratios $(f_{UV}/f_{LyC})_{\rm int}$ need to be modelled for each galaxy individually, where values typically range between 2-10. However, as the intrinsic ratio is sensitive to a host of unknown features within a galaxy's physical properties, such as metallicities, stellar ages and star formation histories (SFHs), we assume a value of $(f_{UV}/f_{LyC})_{int} = 3$.  This is a reasonable average, and allows us to carry out a direct comparison of our results to previous work in the literature where this same assumption is made. This estimate has been shown to be appropriate for a young ($\sim$10$^7$ yr), stellar population of solar metallicity \citep{Rutkowski2016}. We note that low-mass galaxies are expected to be extremely metal poor ($<0.1Z_{\odot}$) such that $(f_{UV}/f_{LyC})_{int} < 3$ \citep[e.g.][]{Bouwens2016,Ceverino2019}. However, it has been shown that the amplitude of the Lyman break does not change considerably down to these metallicities, with intrinsic flux ratios dropping only as low as $\sim$2 \citep[e.g.][]{Leitherer1999,Inoue2005}.  Thus, this assumption does not produce a large uncertainty in our results.

Neutral hydrogen clouds along the line-of-sight towards high redshift galaxies can significantly affect the measured UV fluxes through both continuum and Lyman line absorption. To correct for IGM absorption, denoted by $\tau_{IGM}$, we adopt the \citet{Inoue2014} model constructed using simulations based on observational statistics of the column density and redshift distribution of the Lyman-alpha forest (LAF), Lyman limit systems (LLSs), and damped Lyman-alpha systems (DLAs). 

\begin{figure}
	\includegraphics[width=\columnwidth]{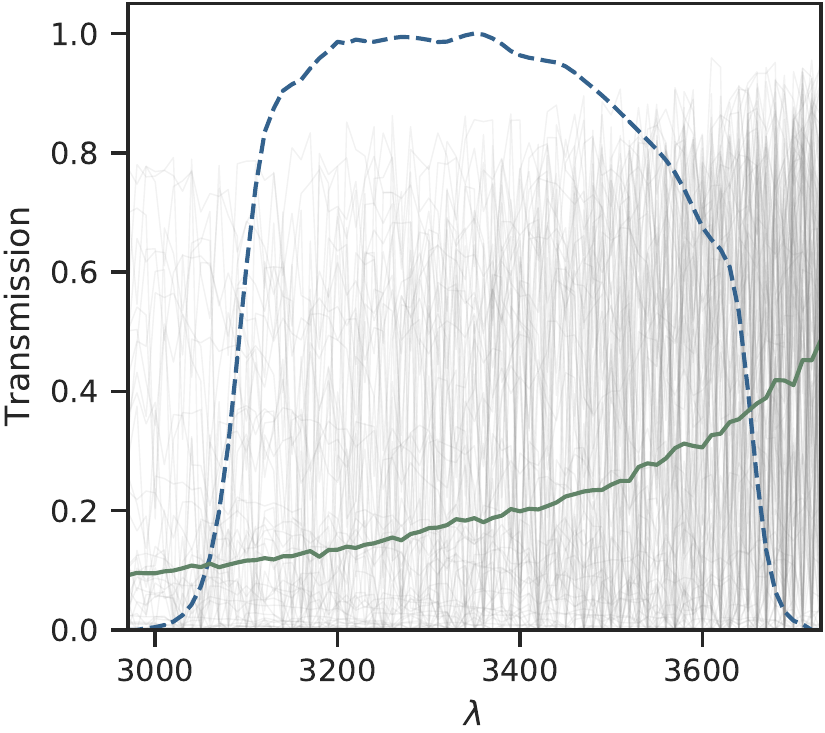}
	\caption{Average IGM transmission of UV light for an example galaxy at $z \approx 3$. Light grey curves show the wavelength dependence of the IGM transmission for one hundred individual runs of our Monte Carlo simulations, which are based on the models of \citet{Inoue2014}. The green curve shows the average IGM transmission of all 1,000 simulated sight lines, while the HST UVIS F336W filter throughput is denoted by the blue dashed line. An estimate for the value of IGM transmission is obtained through the convolution of the corresponding LyC filter response with the average IGM transmission.}
	\label{fig:igm}
\end{figure}

To account for this we carry out a series of Monte Carlo (MC) simulations of IGM absorption, simulating 1,000 lines of sight at the redshift of each candidate as shown in Figure~\ref{fig:igm}. The mean IGM transmission can then be estimated from the resulting probability distribution, convolved with the corresponding LyC filter bandpass.  We also include uncertainties in our photometric redshifts for our galaxies as part of the uncertainty in the IGM optical depth, which we propagate throughout our analysis.  

As explained in \cite{Steidel2018} there may also be some excess line of sight absorption due to transversing the circumgalactic medium (CGM) in the galaxy hosting the LyC emission.   For samples such as ours, at low stellar mass, there has not been a detailed consideration of this effect, and for comparison to other works we do not attempt to correct for it in this paper.    The effect of this would be to only slightly increase our upper limits and detection values, especially given that lower mass systems are likely to contain a smaller CGM.    There are also no galaxies close to the line of sight for our two sources.  Therefore absorption from the CGM of the host or another foreground galaxy is also unlikely to be a major effect.

We thus use these values of IGM transmission along with the observed flux ratio, $(f_{UV}/f_{LyC})_{\rm obs}$, in order to estimate the escape fraction via Equation~\ref{eq:relEF}.

In summary, we find that there are two galaxies in our sample with escaping LyC above 1$\sigma$ when measuring isophotal fluxes as described above.  The properties of these sources are listed in Table~~\ref{tab:results}, along with the other galaxies in our sample at the same redshifts which are not individually detected. The two galaxies with tentative Lyman-continuum detections are numbered 809 and 803 in this table.  For these two galaxies we measure fluxes inside a 0.2\arcsec~radius aperture centered on the peak of the flux in the LyC band. Following a careful quantification of local background levels we find the significance of both of these sources to be $\sim3.2\sigma$. We calculate these galaxy's relative escape fractions as: $f_{esc}^{rel} = 0.032^{+0.081}_{-0.009}$, and 0.021$^{+0.101}_{-0.006}$, respectively.  Both galaxies are low-luminosity systems with M$_{\rm UV} = -18.3$ and $-18.86$, respectively.  They both also have relatively low stellar masses with M$_{*} < 10^{9}$ M$_{\odot}$.  These two systems are also among those at the higher end of our redshift range at $z = 3.46$ and $3.28$, respectively.

For non-detections ($<1\sigma$ within the first examination), we provide upper flux limits (1$\sigma$) based on the background measurements considering only the pixels associated with the isophotes for each galaxy. The properties of these galaxies are summarised in Table~\ref{tab:results} for which we find an average upper limit for each galaxy's escape fraction within our sample.  We find a wide range of upper limits on the escape fraction, depending on the properties of each system, but most have limits at $< 0.07$.  

\subsection{Stacked limits}

In this section we discuss in more detail how we calculate our stacked limits for the Lyman-continuum escape fraction for our sample.  To do this, we create stacked images in both the rest-frame ionizing and the non-ionizing bands for all non-detections in our sample within the same HST/UVIS filter mosaics.   We do not include the two galaxies for which have individual detections when carrying out this stacking.   This means that our limits do not include these two systems and the results must be understood in this content as a limit of detections for systems within our observational set up.

As we are investigating galaxies over the redshifts range of $2 < z < 3.5$, simply stacking all objects in our emission line sample would not provide robust estimates, mainly due to variations in the IGM transmission and galaxy properties at varying redshifts. Thus, we stack candidates based on the rest-frame LyC filter, limiting the redshift range, and minimising variations in $\tau_{IGM}$.  By limiting the redshift range of objects included in the stacks we also reduce contributions from non-ionizing flux in the LyC measurements, and also the red leak in the F275W filter.   We stack the rest-frame ionizing flux in the HST/UVIS F275W and F336W filters, utilising the HST/ACS F606W filter for the corresponding rest-frame UV stacks.   We do these stacks in several different ways, altering our methodology to test any variation in our results.   We also apply this procedure independently for each cluster field. 

\begin{figure*}
	\includegraphics{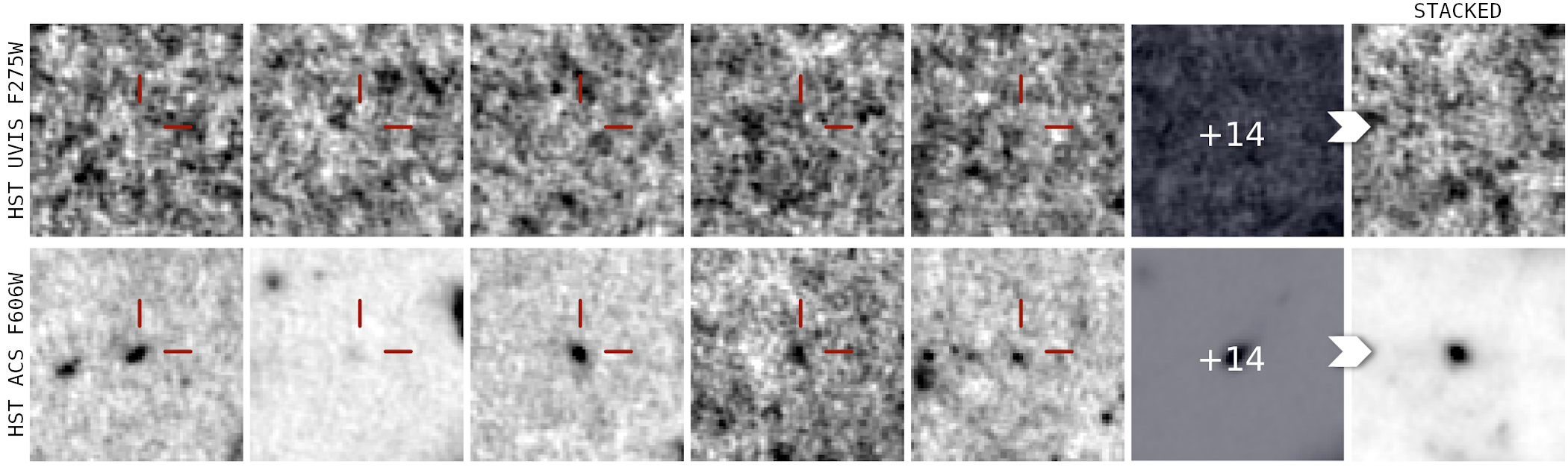}
	\caption{Example postage stamp cut-outs and stacked images for objects in the HST UVIS F275W band in the A0370 cluster. The top row shows the F275W band cut outs representing the LyC at $\lambda_{\textrm{rest}} \leq 912$\AA. HST ACS F606W cut-outs on the bottom row depict the non-ionizing flux at $\lambda_{\textrm{rest}} \approx 1500$\AA. The right most column shows the final stacked image, while all other columns are individual galaxies included in the stack. In total, 19 galaxies are included in the stack.  As described in the text, we investigate various methods of stacking, including different centres and neighbor removal, with one example shown here.} All images are shown prior to cleaning, however the final measurements are taken after the removal of neighbouring objects and contaminants. Postage stamps are 4\arcsec along each side.
	\label{fig:sA275}
\end{figure*}

\begin{figure*}
	\includegraphics{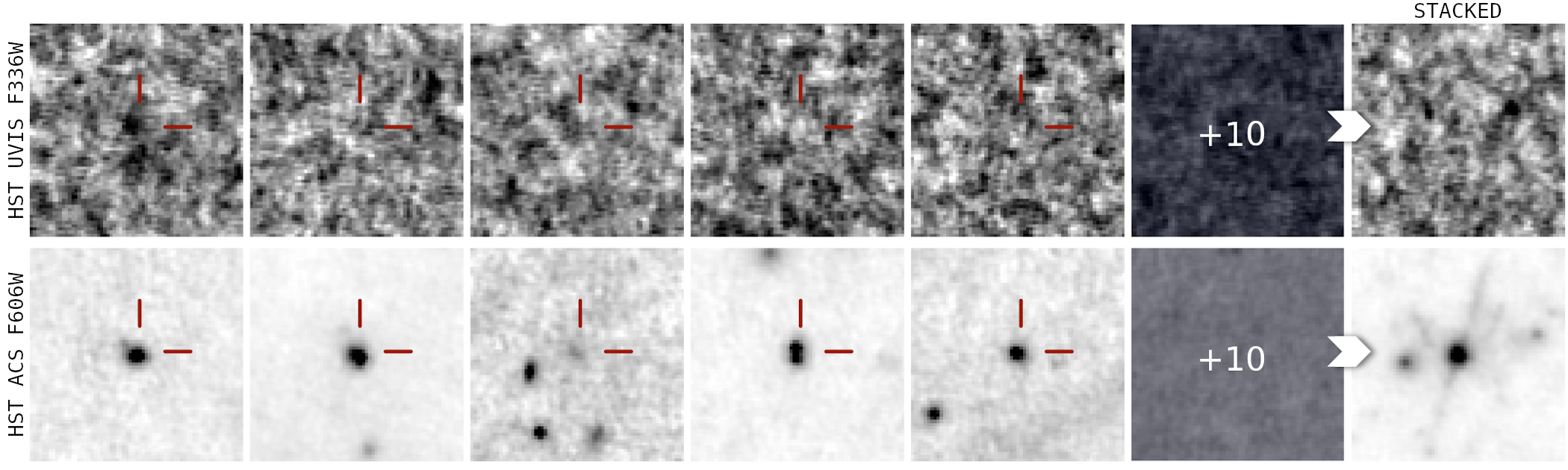}
	\caption{Example postage stamp cut outs and stacked images for objects in the HST UVIS F336W band in the A0370 cluster. This is the same as Figure~\ref{fig:sA275}, however the top row shows the F336W band and a total of 15 galaxies are included in the stack. See the caption to \ref{fig:sA275} and text for details on the stacking methodology, with only one method demonstrated here.}
	\label{fig:sA336}
\end{figure*}

We initially followed a simple stacking procedure, summing all pixels in $4\arcsec \times 4$\arcsec~postage stamps for both the rest-frame LyC and UV filters.  However, no significant detections are seen through this basic stacking method. As described in \citet{Siana2010}, variations in the sizes and morphologies of the sample galaxies can produce unreliable measurements within stacked data. For example, a strong flux measured in a pixel of one postage stamp may be compromised by noise within others where the spatial extent of the stacked galaxies can vary significantly. To account for these issues we perform an optimised stacking procedure.   As detailed in \citet{Siana2010}, we sum only over the pixels in the LyC and UV stamps that are associated with the isophotes defined by the UV band images. This stacking procedure also provides no statistically significant detection in the LyC. Furthermore, we carried out a series of stacking procedures whereby we change slightly the positions of the centers of the stacks, as well as stack the directly observed, as well as demagnified, images. We show visual examples of one of our stacking procedure in Figures~\ref{fig:sA275} and \ref{fig:sA336} for the two A0370 cluster stacks. As can be seen there may be some off-center residuals after the stacking has been carried out depending on how the alignment of the galaxies in question is done.  To avoid this issue, our stacking procedure investigated using slightly different centers to uncover the importance of these offsets in detecting a signal. We ultimately find that the stacking limits do not depend on this, and essentially the same limits are found through all iterations.  When combining the data we also investigate several methods including inverse variance weighting and uniform weighting, finding essentially no difference.  

We also carefully mask out neighboring objects before any stacking is carried out to ensure as much as possible that contamination from other galaxies is not entering the signal when the stack is carried out.  However, this process is not fool-proof as it is possible that not all this light is identified correctly.  In general, however, any contamination from other galaxies would only raise our escape fraction limit.  Since we find already very low limits, this ensures that the value of the typical escape fraction for our sample is lower than this. 

As mentioned in Section~\ref{sec:ef}, the nature of the IGM and the uncertain intrinsic flux ratios of these galaxies make the accurate measurement of average or limits of escape fractions difficult. For this reason we also implement stacking procedures for measuring the limits of detections, enabling the study of properties of the galaxy sample in a more statistically significant manner. Due to the wide redshift range of our sample we stack objects from each cluster into bins based on the available filters, providing 4 stacks in total. We do however note that due to the small sample size, particularly for the candidates identified in the M1149 imaging, stacked results may not be representative of the galaxy population as a whole.

Note, however, that the galaxies in our stacks have a range of lensing magnification factors. This fact slightly complicates how stacking can be done and how to interpret our values.  We stacked our samples both with and without demagnifying each galaxy's light by its magnification factor. This simply means dividing the oberved pixel counts in the area of the galaxy we stack by this magnification factor.  we do not find any significant detection both before and after demagnifying fluxes by the magnification factor. However, to obtain our quantified escape fraction limits, we demagnify the pixel values before stacking.  This essures that the values we stack are true limits and not biased by any possible boosting by magnification.  We present our stacked values in Table~\ref{tab:stackres}.

We also carry out our analysis after extra `cleaning' of the images to ensure our results are reliable.  Using the segmentation maps to isolate neighbouring sources within the postage stamps, we use a modified version of the python package galclean\footnote{\url{https://github.com/astroferreira/galclean}}. As described in \citet{Ferreira2018}, galclean makes use of Astropy's photoutils package \citep{Astropy} to replace contaminating pixels with random values sampled from the noise, such that the background distribution of each image remains unchanged. The images are then stacked as previously described.  When we carry out our measurements again on these cleaned images we also find no flux above background levels in the LyC stacks.

In Figure~\ref{fig:ef}, we show the relative escape fraction ($f_{esc}^{rel}$) as a function of redshift for our sample. These relative escape fractions are corrected for IGM transmission, estimated via our Monte Carlo simulations of 1,000 sight lines per object.  Our tentative detected sources, and the upper limits from Table~\ref{tab:results} are displayed as orange points and small blue arrows respectively, while stacked limits (Table~\ref{tab:stackres}) are shown as solid arrows. Additionally, we include various results from the literature, spanning the entire redshift range of our sample. 

In summary, we place stacked limits for some selections for our sample on the escape fraction to be $< 0.01$ and $< 0.02$ in the Abell 370 field, and slightly higher for the MACS field.   Our stack limits are thus between 1-3\% for the escape fractions for our systems. This implies that, outside of our two direct detections, the escape fractions for these low-mass galaxies with line emission is not high. We discuss the implications of this later in \S 4.1.

\begin{figure*}
    \centering
	\includegraphics[width=\textwidth]{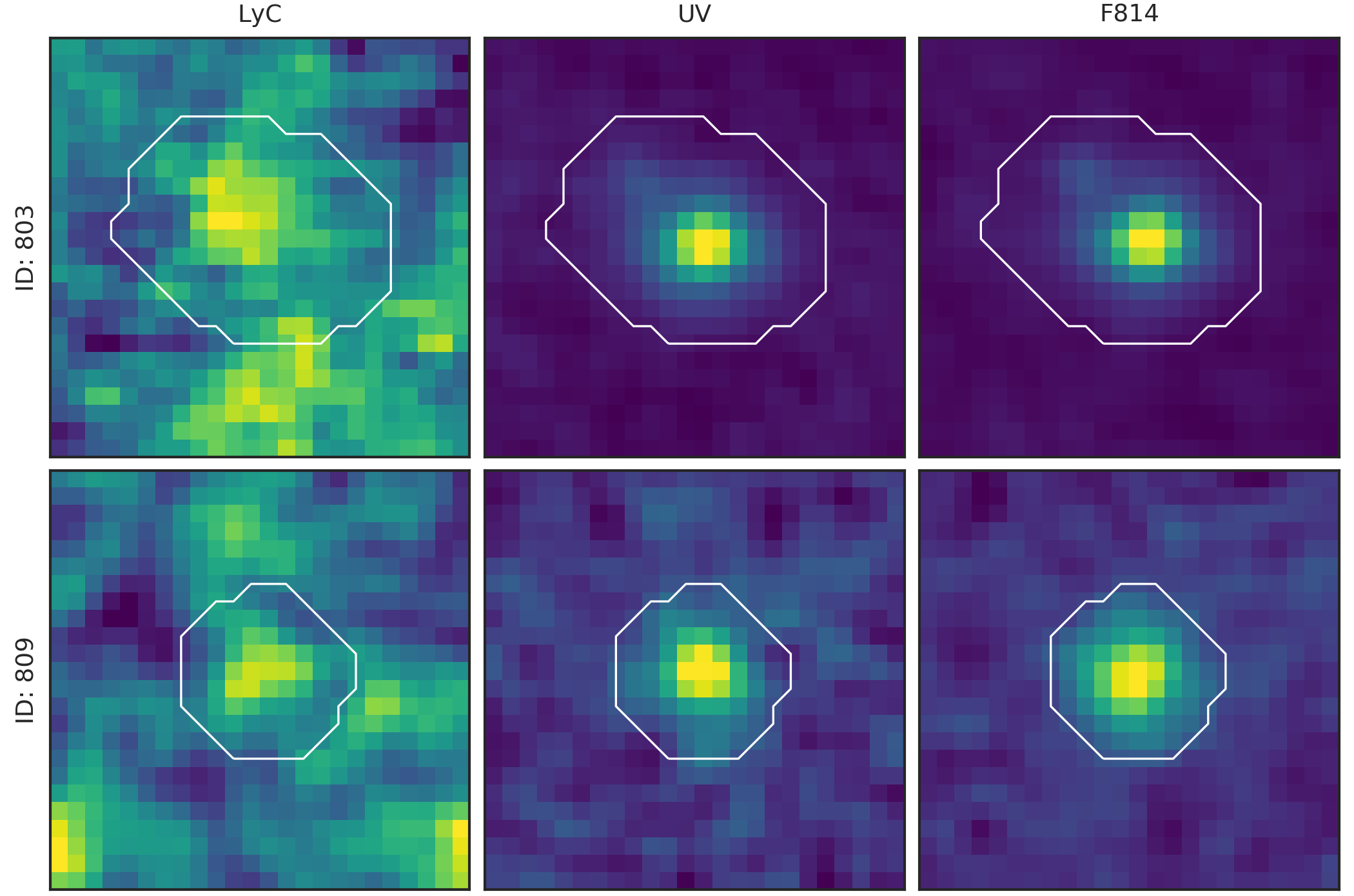}
	\caption{Lyman-continuum (left), UV (middle) and F814W (right) postage stamps for the two galaxies in our sample with flux measurements of $>2\sigma$ in the Lyman-continuum. Postage stamps are $\sim$ 1.5\arcsec $\times$ 1.5\arcsec~in size. Object ID:803 is located in the M1149 cluster and is detected in the F336W band at redshift z=3.28. Source ID:809 is also detected in the F336W band, and is located in the A0370 cluster field at $z=3.28$}
	\label{fig:detections}
\end{figure*}

\begin{table*}
\footnotesize
\caption{Summary of individual escape fraction of emission line selected sample}
\label{tab:results}
\begin{tabular*}{\textwidth}{l@{\extracolsep{\fill}}cclcccrccr}
\hline \\ [-3ex]
ID & R.A. & DEC. & Redshift (z) & M$_{\small\rm{UV}}$ & M$_*$ & $\mu$ & A & {$\tau_{\rm{IGM}}$} & \multicolumn{1}{c}{$f_{esc}^{rel}$} \\
    & [deg] & [deg] & & [mag] & [$\log($M$_*$/M$_{\odot})$] & & & \\
(1) & (2) & (3) & ~~(4) & (5) & (6) & (7) & (8) & (9) & \multicolumn{1}{c}{(10)} \\ [1mm]
\hline \\ [-4mm]
\multicolumn{10}{l}{{\underline{MACS J1149.5+2223}}} \\ [1mm]
F275W:
\vspace{0.5mm}\\
1283 & 177.38723 & 22.38438 & 2.81 $\pm$ 0.04$^*$        & -19.32$\pm$0.08            & 9.40$\pm$0.14 & 1.51  & 0.23 $\pm$ 0.03 & 0.12 $\pm$ 0.08 & $<$ 0.02            \\
2545 & 177.39976 & 22.39301 & 2.80 $\pm$ 0.03$^*$        & -18.00$\pm$0.15            & 8.80$\pm$0.22 & 14.25 & 0.21 $\pm$ 0.02 & 0.15 $\pm$ 0.09 & $<$ 0.02            \\
4749 & 177.41050 & 22.40471 & 2.43 $\pm$ 0.28            & -20.39$\pm$0.05            & 9.50$\pm$0.13  & 1     & 0.43 $\pm$ 0.07 & 0.37 $\pm$ 0.13 & $<$ 0.25            \\
5565 & 177.40976 & 22.41252 & 2.98 $\pm$ 0.20            & -18.15$\pm$0.14            & 9.91$\pm$0.10  & 1.71  & 0.57 $\pm$ 0.01 & 0.12 $\pm$ 0.08 & $<$ 0.01            \\ [1mm]
F336W:
\vspace{0.5mm}\\
809  & 177.38238 & 22.37989 & 3.46 $\pm$ 0.19            & -18.30$\pm$0.13            & 7.81$\pm$0.42  & 1.04  & 0.12 $\pm$ 0.05 & 0.09 $\pm$ 0.08 &   0.032$^{+0.081}_{-0.009}$ \\
4500 & 177.42167 & 22.40612 & 3.48 $\pm$ 0.68            & -16.87$\pm$0.23            & 8.42$\pm$0.30  & 1.40  & 0.31 $\pm$ 0.14 & 0.08 $\pm$ 0.07 & $<$ 0.04            \\
4971 & 177.42151 & 22.40881 & 3.15 $\pm$ 0.05            & -19.81$\pm$0.07            & 8.29$\pm$0.35  & 1.33  & 0.36 $\pm$ 0.02 & 0.23 $\pm$ 0.18 & $<$ 0.02            \\
[1mm] \hline \\ [-4mm]
\multicolumn{10}{l}{{\underline{Abell 370}}} \\ [1mm]
F275W: 
\vspace{0.5mm}\\
1341 & 39.95829 & -1.59271 & 2.73 $\pm$ 0.20             & -16.92$\pm$0.23             & 7.18$\pm$0.55  & 2.81  & 0.30 $\pm$ 0.07 & 0.19 $\pm$ 0.11 & $<$ 0.04            \\
1872 & 39.94373 & -1.58733 & 2.62 $\pm$ 0.54             & -18.53$\pm$0.12             & 8.64$\pm$0.28  & 1     & 0.45 $\pm$ 0.12 & 0.24 $\pm$ 0.12 & $<$ 0.06            \\
2117 & 39.95892 & -1.58499 & 2.95 $\pm$ 0.37             & -16.93$\pm$0.22             & 8.36$\pm$0.33  & 2.89  & 0.39 $\pm$ 0.10 & 0.12 $\pm$ 0.08 & $<$ 0.03            \\
2164 & 39.96706 & -1.58455 & 2.91 $\pm$ 0.04$^{\dagger}$ & -16.54$\pm$0.27             & 7.67$\pm$0.45 & 11.31 & 0.14 $\pm$ 0.03 & 0.13 $\pm$ 0.09 & $<$ 0.03            \\
2548 & 39.98755 & -1.58159 & 2.75 $\pm$ 1.62             & -15.71$\pm$0.37             & 7.54$\pm$0.41  & 2.48  & 0.17 $\pm$ 0.09 & 0.19 $\pm$ 0.11 & $<$ 0.08            \\
3266 & 39.98290 & -1.57647 & 2.57 $\pm$ 1.70             & -15.87$\pm$0.35             & 7.66$\pm$0.38  & 4.87  & -               & 0.27 $\pm$ 0.12 & $<$ 0.45            \\
3706 & 39.99485 & -1.57280 & 2.61 $\pm$ 0.46             & -17.56$\pm$0.17             & 8.58$\pm$0.24 & 1.49  & 0.22 $\pm$ 0.03 & 0.25 $\pm$ 0.12 & $<$ 0.04            \\
3854 & 39.94983 & -1.57175 & 2.66 $\pm$ 0.32             & -18.04$\pm$0.14             & 8.33$\pm$0.32  & 1     & 0.14 $\pm$ 0.04 & 0.23 $\pm$ 0.11 & $<$ 0.03            \\
4114 & 39.99397 & -1.57058 & 2.61 $\pm$ 0.06             & -20.12$\pm$0.06             & 9.12$\pm$0.19  & 1.52  & 0.39 $\pm$ 0.02 & 0.27 $\pm$ 0.12 & $<$ 0.03            \\
4213 & 39.99166 & -1.56925 & 2.63 $\pm$ 0.77             & -16.55$\pm$0.27             & 6.97$\pm$0.55  & 1     & -               & 0.24 $\pm$ 0.12 & $<$ 0.20            \\
4366 & 39.95659 & -1.56795 & 2.45 $\pm$ 0.55             & -15.60$\pm$0.39             & 7.49$\pm$0.52  & 2.98  & 0.18 $\pm$ 0.07 & 0.36 $\pm$ 0.14 & $<$ 0.11            \\
4519 & 39.99003 & -1.56789 & 2.67 $\pm$ 0.04             & -21.05$\pm$0.04             & 9.87$\pm$0.11  & 1     & 0.60 $\pm$ 0.03 & 0.22 $\pm$ 0.11 & $<$ 0.05            \\
4616 & 39.99128 & -1.56780 & 2.85 $\pm$ 0.06             & -19.72$\pm$0.07             & 8.68$\pm$0.24  & 1     & 0.40 $\pm$ 0.09 & 0.15 $\pm$ 0.09 & $<$ 0.05            \\
4989 & 39.96107 & -1.56262 & 2.81 $\pm$ 0.05$^{\dagger}$ & -19.44$\pm$0.08             & 8.29$\pm$0.31  & 5.68  & 0.30 $\pm$ 0.01 & 0.19 $\pm$ 0.11 & $<$ 0.01            \\
5226 & 39.97361 & -1.55974 & 3.00 $\pm$ 0.07             & -19.49$\pm$0.08             & 10.23$\pm$0.12 & 1     & 0.36 $\pm$ 0.04 & 0.10 $\pm$ 0.07 & $<$ 0.01            \\
5347 & 39.95198 & -1.56308 & 2.44 $\pm$ 0.35             & -20.64$\pm$0.05             & 9.84$\pm$0.15  & 2.26  & 0.57 $\pm$ 0.09 & 0.37 $\pm$ 0.14 & $<$ 0.07            \\
5687 & 39.97501 & -1.55618 & 2.82 $\pm$ 0.03             & -21.21$\pm$0.03             & 10.19$\pm$0.12 & 1     & 0.07 $\pm$ 0.01 & 0.16 $\pm$ 0.10 & $<$ 0.02            \\
5769 & 39.97072 & -1.55482 & 2.53 $\pm$ 0.67             & -17.05$\pm$0.22             & 8.50$\pm$0.25  & 1     & 0.33 $\pm$ 0.04 & 0.30 $\pm$ 0.13 & $<$ 0.08            \\[1mm]
F336W:
\vspace{0.5mm}\\
803  & 39.97475 & -1.59918 & 3.28 $\pm$ 0.06$^{\dagger}$ & -18.86$\pm$0.10             & 8.91$\pm$0.22  & 3.43  & 0.23 $\pm$ 0.02 & 0.21 $\pm$ 0.17 &   0.021$^{+0.101}_{-0.006}$ \\
1399 & 39.98341 & -1.59220 & 3.01 $\pm$ 0.09             & -18.13$\pm$0.14             & 7.41$\pm$0.45  & 3.19  & 0.79 $\pm$ 0.04 & 0.32 $\pm$ 0.20 & $<$ 0.05            \\
1629 & 39.95825 & -1.59015 & 3.23 $\pm$ 0.06             & -19.14$\pm$0.09             & 9.52$\pm$0.14  & 2.64  & 0.50 $\pm$ 0.03 & 0.17 $\pm$ 0.15 & $<$ 0.04            \\
2068 & 39.97426 & -1.58559 & 3.13 $\pm$ 0.07$^{\dagger}$ & -17.62$\pm$0.17             & 7.66$\pm$0.45 & 6.76  & 0.34 $\pm$ 0.04 & 0.26 $\pm$ 0.19 & $<$ 0.04            \\
2128 & 39.95285 & -1.58534 & 3.24 $\pm$ 0.05             & -20.16$\pm$0.06             & 9.68$\pm$0.15  & 1.99  & 0.30 $\pm$ 0.01 & 0.16 $\pm$ 0.09 & $<$ 0.01            \\
2282 & 39.96371 & -1.58327 & 3.39 $\pm$ 0.61             & -16.51$\pm$0.27             & 8.94$\pm$0.19 & 1     & 0.58 $\pm$ 0.10 & 0.11 $\pm$ 0.07 & $<$ 0.06            \\
2330 & 39.96161 & -1.58302 & 3.81 $\pm$ 1.14$^{\dagger}$ & -15.25$\pm$0.44             & 7.15$\pm$0.51  & 4.45  & 0.38 $\pm$ 0.15 & 0.27 $\pm$ 0.12 & $<$ 0.16            \\
2446 & 39.95052 & -1.58232 & 3.06 $\pm$ 1.49             & -16.76$\pm$0.25             & 8.32$\pm$0.27  & 1     & 0.47 $\pm$ 0.14 & 0.27 $\pm$ 0.19 & $<$ 0.29            \\
2558 & 39.95363 & -1.58185 & 2.92 $\pm$ 0.05$^{\dagger}$ & -20.28$\pm$0.05             & 9.08$\pm$0.17  & 2.16  & 0.36 $\pm$ 0.06 & 0.29 $\pm$ 0.20 & $<$ 0.02            \\
2997 & 39.96986 & -1.58070 & 2.75 $\pm$ 0.04$^{\dagger}$ & -18.14$\pm$0.14             & 8.07$\pm$0.31  & 19.48 & 0.35 $\pm$ 0.07 & 0.24 $\pm$ 0.18 & $<$ 0.02            \\
3023 & 39.95382 & -1.57837 & 3.19 $\pm$ 0.13             & -18.01$\pm$0.15             & 8.84$\pm$0.23  & 2.46  & 0.60 $\pm$ 0.10 & 0.19 $\pm$ 0.16 & $<$ 0.02            \\
3721 & 39.99461 & -1.57270 & 3.18 $\pm$ 1.02             & -17.60$\pm$0.17             & 8.41$\pm$0.29  & 1.48  & 0.42 $\pm$ 0.09 & 0.19 $\pm$ 0.13 & $<$ 0.08            \\
4113 & 39.95380 & -1.57002 & 3.17 $\pm$ 0.13             & -18.94$\pm$0.10             & 9.39$\pm$0.16  & 2.69  & 0.35 $\pm$ 0.02 & 0.19 $\pm$ 0.16 & $<$ 0.02            \\
4346 & 39.99053 & -1.56810 & 3.17 $\pm$ 0.29             & -18.73$\pm$0.11             & 9.31$\pm$0.15  & 1     & 0.63 $\pm$ 0.11 & 0.20 $\pm$ 0.14 & $<$ 0.10            \\
4620 & 39.99227 & -1.56928 & 3.49 $\pm$ 0.06             & -20.67$\pm$0.04             & 10.02$\pm$0.08 & 1     & 0.15 $\pm$ 0.01 & 0.08 $\pm$ 0.07 & $<$ 0.01            \\ 
4797 & 39.99221 & -1.56393 & 3.07 $\pm$ 0.14             & -18.90$\pm$0.10             & 9.31$\pm$0.15 & 1     & 0.13 $\pm$ 0.02 & 0.26 $\pm$ 0.17 & $<$ 0.04            \\
5157 & 39.96892 & -1.56041 & 3.22 $\pm$ 1.22             & -17.20$\pm$0.20             & 7.30$\pm$0.47 & 1     & 0.05 $\pm$ 0.01 & 0.18 $\pm$ 0.15 & $<$ 0.09            \\
[1mm] \hline \\ [-2mm]
\end{tabular*}
\begin{minipage}[c]{\textwidth}
\textbf{Columns:} (1) HFF catalogue ID; (2) Right ascension in degrees; (3) Declination in degrees; (4) Photometric redshift estimated using the EAZY software with 1$\sigma$ confidence intervals computed from the $p(z)$ probability distribution. This includes all available SHARDS and HFF-DeepSpace bands.  Spectroscopic redshifts are used when available in which $*$ and $\dagger$ denote values obtained from HFF-DeepSpace and MUSE catalogs \citep{Lagattuta2019} respectively; (5) Magnification corrected isophotal magnitude in the rest-frame UV ($\lambda_{\textrm{rest}} \approx 1500$\AA), HST/ACS F606W is used for all objects; (6) Magnification corrected stellar mass in units of log solar masses, estimated with FAST utilising all available bands. Typical errors are 0.2 dex in log units; (7) Median magnification factor ($\mu$) from all available lens models, when data is unavailable we set $\mu$ to 1 ; (8) Asymmetry value used for merger classification; (9) Average IGM transmission of UV light based on Monte Carlo simulations over 1,000 simulated sight lines; (10) The relative LyC escape fraction and its $\pm$ 1$\sigma$ uncertainties, including corrections for dust and IGM attenuation. \textbf{Note:} Where escape fractions are preceded with an `$<$', the values represent 1$\sigma$ upper limits.
\end{minipage}
\end{table*}

\begin{table*}
\caption{Summary of escape fractions for stacked images}
\label{tab:stackres}
\centering
\begin{tabular*}{\textwidth}{@{\extracolsep{\fill}}lcccrcccr}
\hline \\ [-2ex]
Cluster & Filter & $N_{obj}$ & {z}  &  \multicolumn{1}{c}{M$_{\rm{UV}}$} & {M$_*$} & {$\mu$} & {$\tau_{\rm{IGM}}$} & \multicolumn{1}{c}{$f_{esc}^{rel}$} \\
 & &  &  &  \multicolumn{1}{c}{[mag]} & [$\log($M$_*$/M$_{\odot})$] & & & \\
(1) & (2) & (3) & (4) & \multicolumn{1}{c}{(5)} & (6) & (7) & (8) & \multicolumn{1}{c}{(9)} \\ [1mm]
\hline \\ [-2mm]
MACS J1149.5+2223 & F275W & 4  & 2.76 & -18.97 & 9.40 & 4.62 & 0.19 & $<$ 0.04 \\
MACS J1149.5+2223 & F336W & 2  & 3.36 & -18.34 & 8.18 & 1.37 & 0.13 & $<$ 0.02 \\
Abell 370         & F275W & 18 & 2.75 & -18.01 & 8.44 & 2.67 & 0.22 & $<$ 0.02 \\
Abell 370         & F336W & 15 & 3.14 & -18.45 & 8.76 & 3.26 & 0.21 & $<$ 0.01 \\ 
[1mm] \hline \\ [-2mm]
\end{tabular*}
\begin{minipage}[c]{\textwidth}
\textbf{Table columns:} (1) Cluster field from which objects have been identified; (2) Filter used for stacking; (3) Number of galaxies in the stack; (4-8) Same as Table~\ref{tab:results}, values are calculated as the mean of all corresponding stacked objects; (9) the 1$\sigma$ upper limit on relative LyC escape fraction.
\end{minipage}
\end{table*}

\begin{figure*}
	\includegraphics[width=\textwidth]{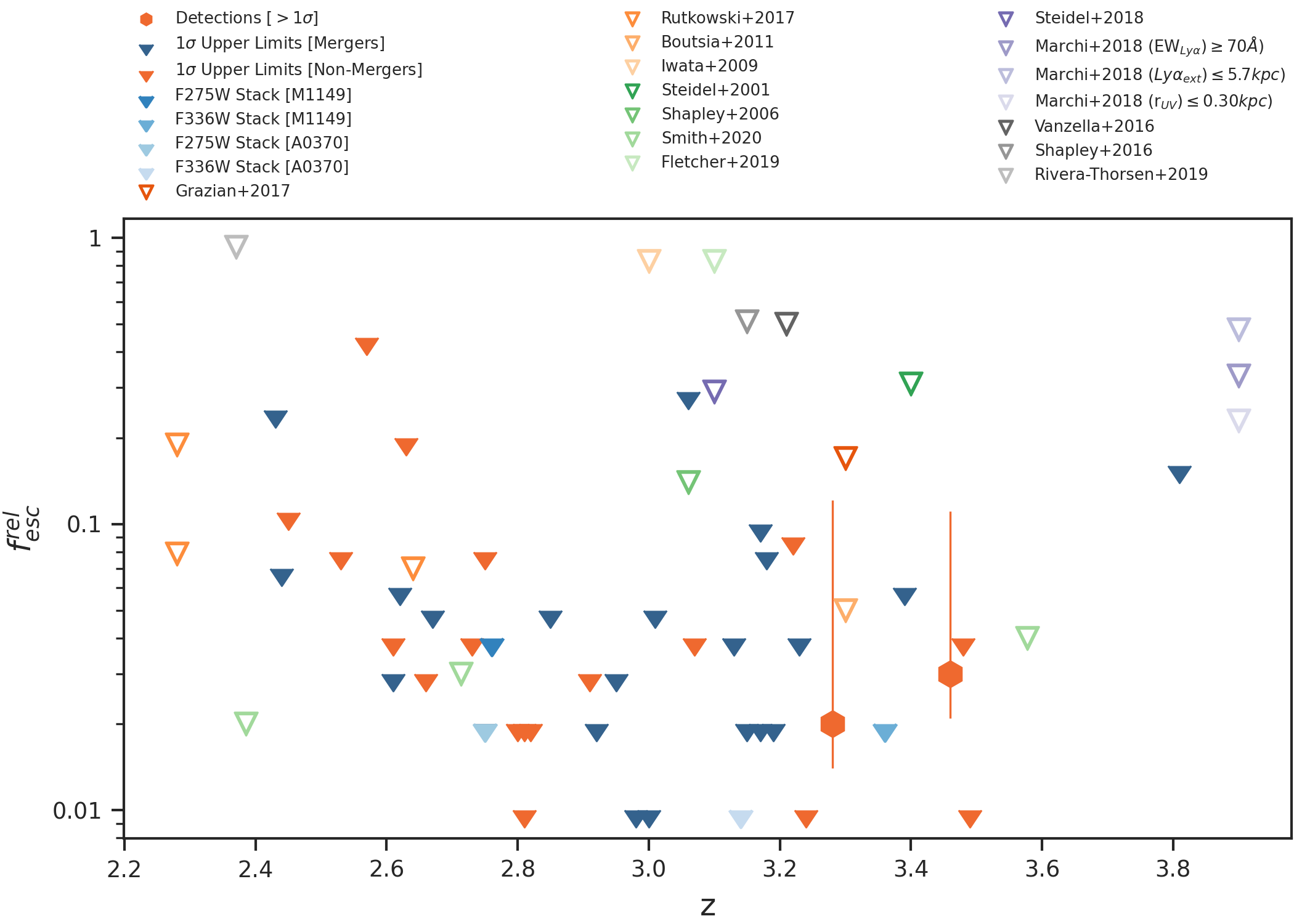}
        \includegraphics[width=\textwidth]{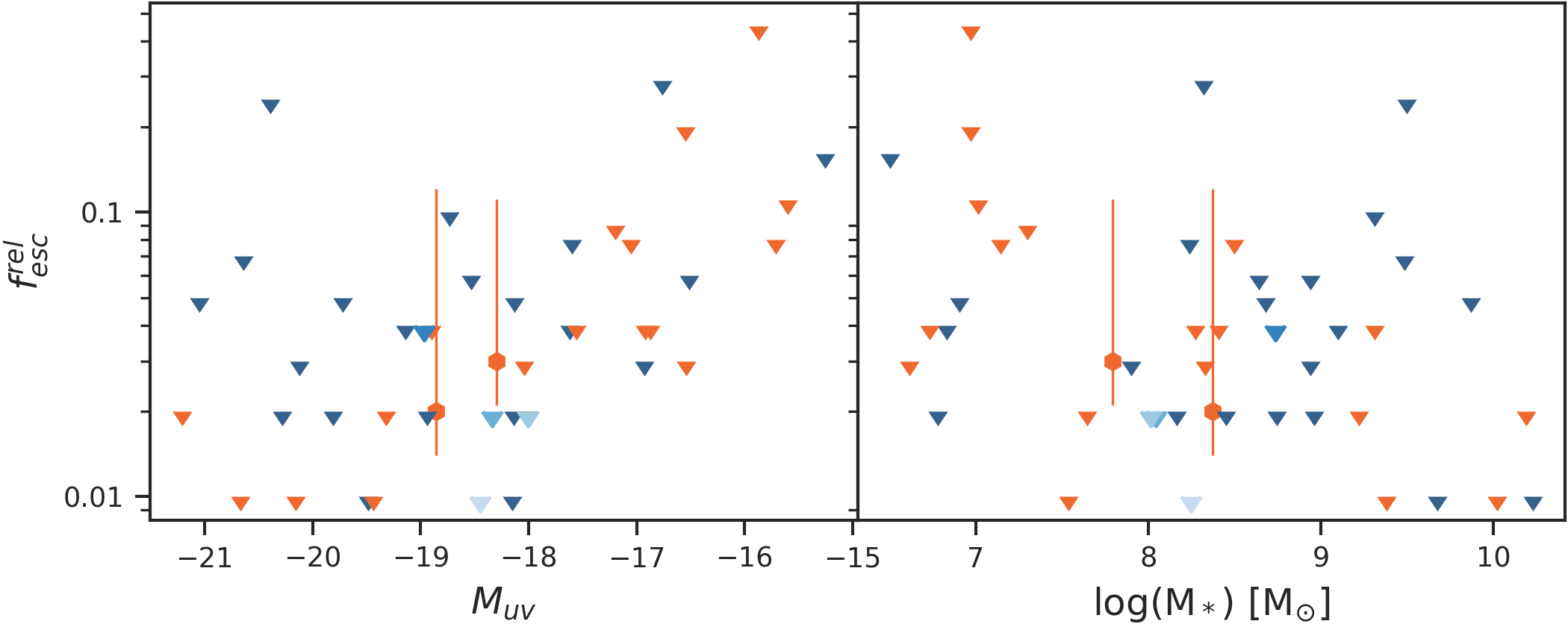}
	\caption{Relative escape fractions as a function of redshift (top), M$_{\rm{UV}}$ (bottom left) and demagnified stellar mass (bottom right). The orange points represent individual galaxies in which we have obtained flux measurements of more than 1$\sigma$ significance in the LyC images (tentative detections), while 1$\sigma$ upper limits for the remainder of the sample are shown by small blue and orange arrows (which are non-detections) for high-asymmetry ($A \geq 0.35$) and low-asymmetry ($A < 0.35$) systems. The properties of these individual measurements are detailed in Table~\ref{tab:results}. Large filled arrows show the results of the stacking procedures as detailed in Table~\ref{tab:stackres}. Empty arrows show stacked results from the literature spanning the entire redshift range of our sample \citep{Steidel2001,Shapley2006,Iwata2009,Boutsia2011,Grazian2017,Rutkowski2017,Smith2020,fletcher2019,Steidel2018,marchi2018}.}
	\label{fig:ef}
\end{figure*}

\subsection{Galaxy Structure}
\label{subsec:asymmetry}

One of our goals is understanding whether galaxy structure correlates with the escape of ionizing radiation. This is a reasonable hypothesis as merging or other dynamically unstable events will disrupt the structure of a galaxy and potentially allow Lyman-continuum photons to escape.  We use the asymmetry parameter \citep{Conselice2000} $A$, from the CAS system \citep{Conselice2003} to quantify the structures of our sources. This is one of the most robust non-parametric morphology measurement methods and is often used as a way to estimate merger fractions based on a galaxy's morphology \citep{Conselice2003a, Conselice2008, bluck2012}.  Note that all our morphological measurements are carried out on the reddest HST filters.

We focus on structure quantification using the asymmetry ($A$) index, as it captures the most important structural features at this resolution and signal to noise level. The asymmetry of a galaxy image is defined as:

\begin{equation}\label{eq:a_definition}
A = \frac{\Sigma|I-I_{180}|}{\Sigma|I|} - A_{BG},
\end{equation}

\noindent where $I$ is the source flux, $I_{180}$ the source flux rotated by $180^\circ$ around its centre and $A_{BG}$ accounts for the asymmetry of the background. The background term, $A_{BG}$, is measured in the same way as the asymmetry but with a patch of the sky that does not overlap with the source's segmentation map.

Additionally, we measure the Concentration index ($C$), as the ratio of two radii each containing a fraction of light from the light from the source as:
\begin{equation}\label{eq:c_definition}
C = 5 \log_{10} \left (\frac{R_{80}}{R_{20}} \right),
\end{equation}
where $R_{80}$ and $R_{20}$ are radii containing $80\%$ and $20\%$ of the flux of the galaxy, respectively. The concentration measurement is a robust indicator of how steep the light profile of the source is, and it is known to correlate with morphological type \citep{Conselice2003}. The space of parameters formed by the asymmetry ($A$) and concentration ($C$) can be then used to identify if a galaxy is potentially a merging system, a late-type galaxy, or an early-type galaxy. 

In the context of mergers, the asymmetry measurement is combined with the smoothness \citep{Conselice2003}, $S$, to define a region of the parameter space dominated by galaxy major mergers using the following criteria:

\begin{equation}\label{eq:asy_cut}
    (A > 0.35), \quad (A > S).
\end{equation}

\noindent The value of $A$ captures how asymmetric the morphology of a galaxy is, and $S$ measures the contribution from high-spatial frequency structures to the flux, such as star-forming regions. The $A > S$ limit ensures that galaxies with pockets of clumpy star formation are not responsible for the bulk asymmetric structures.  

\begin{figure*}
    \centering
    \includegraphics[width=\textwidth]{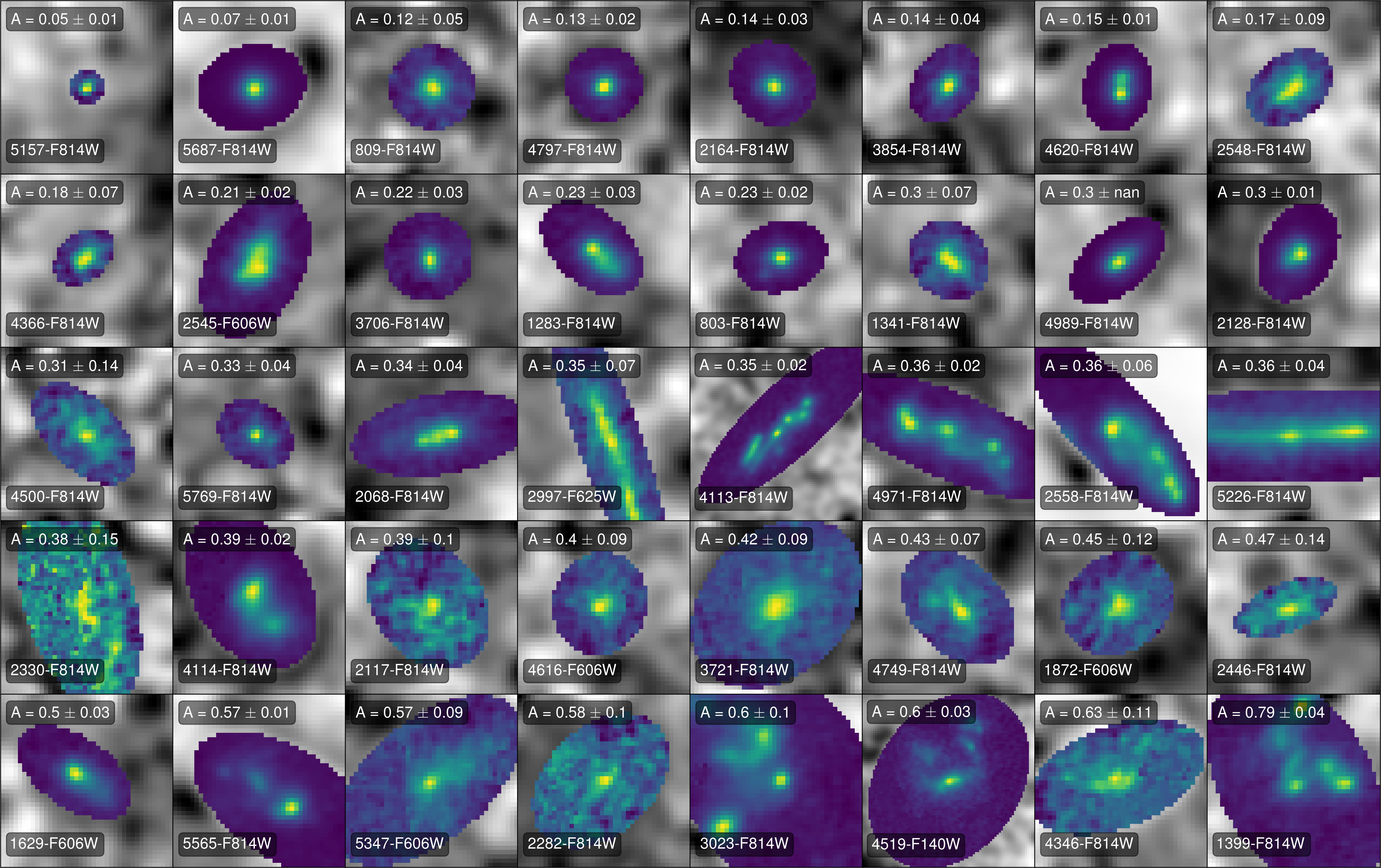}
    \caption{Mosaic showing 40 out of our 42 sources with valid asymmetry measurements listed with their ID and asymmetry values.  Sources 3266 and 4213 are not included due to significant distortions due to magnification. The asymmetry value for each galaxy is located in the upper left corner, while the source identification and band used for the measurement is located at the bottom left corner. Each source image shows the Petrosian region ($1.5 \ R_p$) used to measure the main term of Equation~\ref{eq:a_definition} in each viridis colourmap. The grey area was smoothed with a gaussian kernel for visualisation purposes.  The background term of the asymmetry, $A_{BG}$, is measured with the original pixel values that are beyond the Petrosian region away from any other galaxy.   The values of the magnification for each galaxy are listed in Table~\ref{tab:results}.}
    \label{fig:a_mosaic}
\end{figure*}

To measure the asymmetry we first select the band with the highest signal-to-noise ratio for each source. This means that most measurements are done in the F606W or F814W bands, but the values are consistent with F160W. We generate a segmentation map with the code {\em galclean} and replace neighbouring sources with a sampling of the background to ensure that our $A$ measurements are not affected by it. From this, we measure the Petrosian radius \citep{Petrosian}, ($R_{p}$), and axis ratio, ($q$), using the code {\rm Morfometryka} \citep{Morfometryka} and proceed to estimate the asymmetry of the source by measuring it within an elliptical aperture of $1.5 Rp$, and axis ratio, $q$. Our results are presented in Table~\ref{tab:results}, and a visualisation of the morphologies are shown in Figure~\ref{fig:a_mosaic}, where the $A$ value is printed in the upper left corner of each image, whilst the source identification and band used for the measurement is shown in the bottom left corner.

Accurate segmentation maps and Petrosian regions for the unstacked images of sources 3266 and 4213 are impossible to create accurately due to their low SNR. Additionally, sources 2997 and 5226 contain lensed morphologies which giving artificially high $A$ values. Finally, due to the proximity to bright sources, the source masking process for source 2330 is not ideal and can introduce biases in the $A$ measurement. As such, we have accurate measurements and merger classifications for 40 sources.   

Distortions due to magnification by the clusters can in principle create higher asymmetries, even if they are not readily visible, as in sources 2997 and 5226.   We investigate this and determine that there is no correlation between asymmetry and magnification.  Thus, it is very likely that the galaxies with a high asymmetry are due to the intrinsic structure and not from magnification issues.    Furthermore, we do not find a trend between magnification and escape fraction limit.  This would be expected, in any case, if galaxies are randomly magnified behind the cluster.  

We consider all galaxies above the asymmetry threshold of Equation~\ref{eq:asy_cut} to be highly distorted. We thus find a high asymmetry fraction of $~45\% \pm 11\%$ for our sample.  This is significantly higher than for larger mass galaxies at similar redshifts from previous studies \citep[e.g.,][]{ConseliceArnold2009, Conselice2014}. This is potentially due to our probing lower mass galaxies that are likely more asymmetric than higher mass systems at the same redshifts.   In general, lower mass galaxies in the nearby universe have a high asymmetry, but this light is clumpy due to star forming regions, and thus not due to a dynamical process which alters the bulk structure of the system e.g., \cite{Conselice2003}. We discuss the implications of these measurements in \S 4.2.

\section{Discussion}
\label{sec:discussion}

\subsection{Comparison to Previous Work}

Here we investigate our samples escape fraction properties and compare these with other studies. First, we note that \citet{Xu2016}, \citet{Anderson2017} and \citet{atek2022} find that low-mass galaxies are potentially the dominant contributors to reionization with escape fractions that are anti-correlated with mass and M$_{UV}$. \citet{Xu2016} predict escape fractions that increase from $\sim$5\% in the mass range of 10$^8$-10$^9$ M$_{\odot}$ up to 40-60\% for fainter even lower-mass galaxies, while \citet{Anderson2017} find a power-law dependence of $f_{esc}$ on UV magnitude.  Individually, \citet{Vanzella2016} find that a single star forming galaxy at $z = 3.2$ has an escape fraction which is likely $> 50$\% (see also \citet{Shapley2016} for a similar object.)  A multiply lensed galaxy, dubbed "The Sunburst Arc" is found to have an escape fraction which varies between 14\ -\ 64\% amongst the lensed images at $z = 2.37$ \cite{RiveraThorsen2019}.  These objects overlap with our sample and show that some star forming galaxies can have a high escape fraction. 

We show our limits and measurements of the escape fraction with galaxy mass and UV magnitude in the bottom two panels of Figure~\ref{fig:ef}, and the results in Table~\ref{tab:results}.  Although we have limited data and only upper limits, these results tend to agree to some degree with the simulations of \citet{Hassan2016} and \citet{Yoshiura2017} who find that the escape fractions have relatively low values ($f_{\rm esc} < 15\%$), independent of galaxy mass or magnitude.

While methods for estimating the escape fraction of a galaxy are fairly consistent throughout the literature, the selection of galaxy types is varied. Searches for escaping Lyman-continuum such as those of \citet{Shapley2016}, \citet{Grazian2017}, and \citet{Saxena2022} examine selected star forming galaxies discovered typically in HST observations combined with ground-based data. Similarly, \citet{Steidel2001}, \citet{Iwata2009} and \citet{Boutsia2011} search for combinations of Lyman break galaxies (LBGs) and Lyman-alpha emitters (LAEs) over various field and protocluster environments. Finally, studies such as \citet{Rutkowski2017} and \citet{Smith2020} use pre-selection of candidates via emission line and spectroscopic redshift catalogues. 

Care needs to be taken when comparing our results to previous work, as very little research has been done on measurements of escape fractions for these types of purely emission line selected galaxies at these redshifts which have low masses. Typically, the relatively small fields covered by lensing clusters, where these galaxies can be found, have not provided the sample numbers required for these types of studies. Furthermore, obtaining samples of emission lines galaxies with low masses is complicated by the lack of spectroscopy for such faint systems. 

Individual limits are not as robust as they could be due to uncertainties in the IGM opacity along an individual galaxy's sight lines, and without spectroscopic follow up intrinsic flux ratios are difficult to estimate. By implementing an intrinsic flux ratio of 3 we are able to compare our results to previous work, but note that galaxies of lower mass are expected to be extremely metal poor ($<0.1Z_{\odot}$) such that $2 < (f_{UV}/f_{LyC})_{\rm int} < 3$. 

Our tentative detection of a measurable escape fraction for two of our low-mass systems (M$_*$ = 10$^{7.81}$ M$_{\odot}$ and 10$^{8.91}$ M$_{\odot}$) allows us to investigate the likelihood that low-mass emission line selected galaxies are able to significantly contribute to the reionization of the universe. One way to make this measurement from our data is to consider the amount of Lyman-continuum emission coming from these systems and using standard assumptions about the structure of the high-redshift IGM \citep[e.g.,][]{Duncan2015}. An average escape fraction of $\sim >$0.03 would allow reionization to occur \citep[][]{Finkelstein2019} if a relatively high number of low-mass galaxies are present at high redshift, which indeed seems likely to be the case \citep[e.g.,][]{Duncan2019, conselice2016, Bhatawdekar2019}.  However, an average escape fraction of $\sim 0.03$ can be ruled out for the line emitters we study here, where our stacked limit is lower than this value. This implies that perhaps low-mass, relatively dust free, star forming galaxies may not be the dominant source of reionization, and more detailed study of other low-mass galaxies and their escape fractions are desperately needed to test our results. 

This suggests that line emitting low-mass galaxies at $z < 3.5$ would not emit enough LyC light to reionize the universe, unless significant evolution in the escape fraction for these objects occurs \citep[][]{Finkelstein2019}.   There may also be examples of low mass galaxies that are able to emit significant Lyman continuum radiation \citep[e.g.,][]{marqueschaves2021,flury2022, Vanzella2022}, yet we are not able to find systems such as these in our higher redshift low mass sample, and overall average escape fractions are small \citep[][]{Ji2020}.

The upper limits of LyC detections as derived from our stacking procedure suggest that our sample of low-mass galaxies may have overall similar escape fractions as samples of LBGs, LAEs and SFGs \citep{Steidel2001,Shapley2006,Iwata2009,Boutsia2011,Grazian2017,Rutkowski2017,Smith2020,fletcher2019,Steidel2018,marchi2018, atek2022} as displayed in Figure~\ref{fig:ef}. We note however that our sample is small, for example the largest stack contains only 19 objects, thus, a significantly larger sample is required in order to provide more robust constraints on the possible contribution of this line emitting galaxy population to the ionizing background. However, these systems do not emit copious LyC radiation.  Overall, however it does not appear that line emitting systems are good locations for finding escape Lyman-continuum radiation. This agrees with the findings of \citet{Wang2021} who find the fraction of low redshift Lyman-continuum leakers decreases at a higher level of [SII] emission.

We, however, can make a comparison with the various detections we have already, although these depend strongly on the depth of the data.   We find 2 out of 42 galaxies with LyC, giving a galaxy LyC emitting fraction of 5$\pm3$\%.  In comparison, \citep[]{Steidel2018} detect 15 galaxies out of 124 Lyman-break selected systems, for a fraction of 12$\pm$3\%, which is more than twice as high as our detection fraction. However, we find that the fraction of our sample detected is similar to the fraction of detected emitters for a general star forming galaxy sample at similar redshifts which has a detection fraction of 6$\pm$2\% \citep[][]{Saxena2022}.  It therefore appears that these systems are not copious emitters of LyC emissions.  Our observations place the best limits yet on the possible contribution of analog low-mass line emitter galaxies to the process of reionization.

\subsection{Correlation with Galaxy Structure}
\label{subsec:role_of_mergers}

The emerging theoretical picture of LyC escape is that feedback carves holes through the ISM \citep[e.g.,][and references therein]{Ma2016,Trebitsch2017}. These chimneys of very low {H}{I} column densities are the channels through which the LyC photons can escape from galaxies.   AGN feedback may also form these holes by producing powerful outflows as observed in more massive galaxies \citep{Genzel2014}.  This feedback, and changes in galaxy structure, can be perhaps further enhanced by dynamical events that are creating asymmetries in galaxies.

As mergers are more common at higher redshifts, even back to the epoch of reionization \citep[e.g.][]{Duncan2019}, this is potentially a major way in which Lyman-continuum emission is facilitated to escape from distant galaxies.   During the merger process a galaxy's internal {H}{I} reservoirs can be sufficiently disturbed such that photons which are usually attenuated by the IGM are able to stream freely outside of the galaxy through these low column density chimneys.   Additionally, dynamical events, such as mergers, can lead to an increase of gas around the AGN and induce bursts of star formation, both of which can lead to the increased production of ionizing photons. Whilst there is little observational work on this in the literature, our sample, which benefits from strong gravitational lensing from massive galaxy clusters, in principle provides a way to investigate these effects within fainter/lower-mass galaxies behind these clusters due to lensing magnification.

\begin{figure}
    \centering
    \includegraphics[width=\columnwidth]{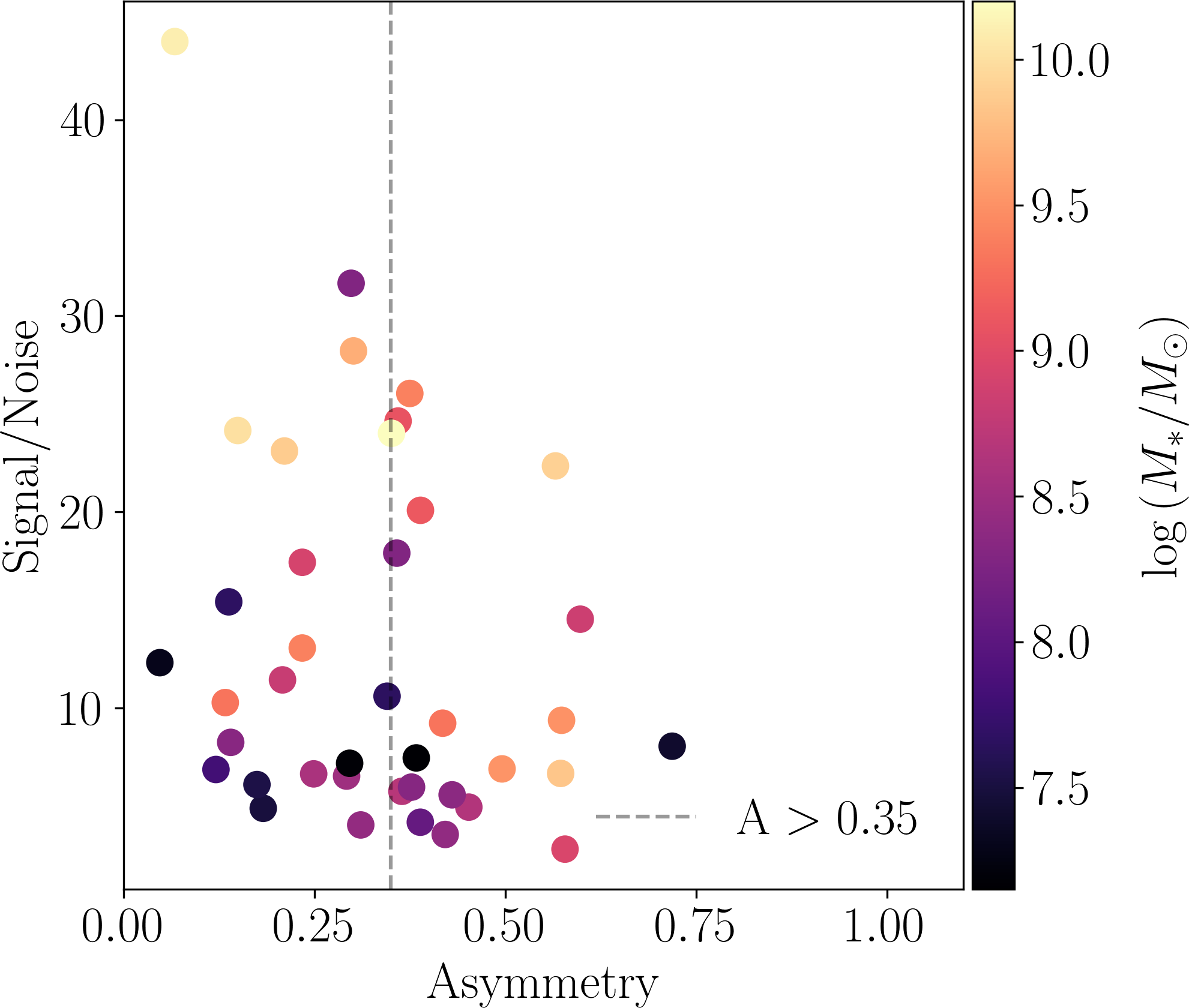}
    \caption{Asymmetry measurements for the galaxies detailed in Section~\ref{subsec:asymmetry}, compared with signal-to-noise on the y-axis, and colour coded by stellar mass. The dashed gray vertical line displays the threshold used to classify a source as a potential merger following Equation~\ref{eq:asy_cut}. The inner plot displays the average $f_{esc}$ limit for galaxy mergers ($A > 0.35$) and non-mergers ($A \le 0.35)$. This separation provides hints at a higher average $f_{esc}$ limit ($7 \pm 2\%$) for galaxies that are likely to be galaxy mergers, as opposed to non-mergers ($4 \pm 1\%$).}
    \label{fig:asy_f}
\end{figure}

\begin{figure}
    \centering
    \includegraphics[width=\columnwidth]{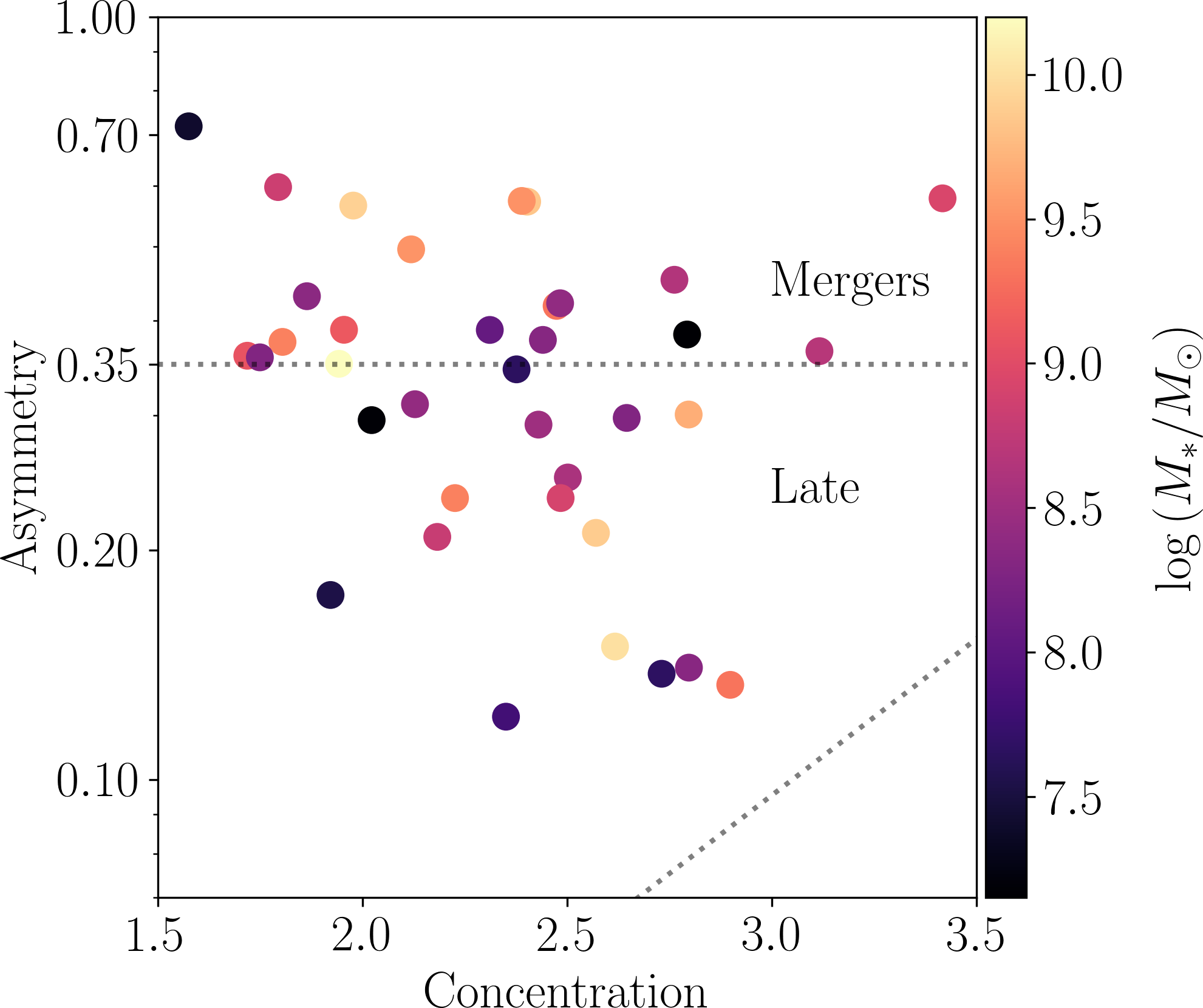}
    \caption{Concentration vs. asymmetry diagram for our sample of line emitters with HST morphologies. Lines denoting the typical regions where mergers and late-type galaxies are located at $z = 0$ based on \cite{Bershady2000} and \cite{Conselice2003}. As can be seen, the sample is mostly divided in merger candidates and late-type galaxies.}
    \label{fig:AxC}
\end{figure}

To carry out this test we separate sources into distorted and non-distorted sub-samples using the asymmetries measured in Section~\ref{subsec:asymmetry}, with the imposed limit from Equation~\ref{eq:asy_cut}.   Figure~\ref{fig:asy_f} shows the asymmetry values for each of our sources as a function of signal-to-noise ratio, and colour coded by stellar mass.  While Figure~\ref{fig:AxC} shows the distribution of our sample in the asymmetry-concentration plane e.g., \cite{Conselice2003}.   We find that $~45\% \pm 11\%$ of the candidates have high asymmetries, often interpreted as a sign of merging, based on their asymmetry  values.   As the CAS system for finding mergers has not been calibrated at such high redshifts and low masses, we do not claim that these are necessarily mergers, and prefer to describe them as highly asymmetric. Thus, our sample is divided almost equally between asymmetric galaxies and those that are not asymmetric. As before, there is no strong evidence that lensing is systematically affecting these structures unequally between these two selected samples.  Overall, we calculate that asymmetric galaxies have a stacked escape fraction limit of $<7\%$, while the more symmetric galaxies have a limit of $<4\%$.   These limits are similar and thus we cannot conclude that distorted structures are statistically more likely to have a higher LyC limit.  

 We cannot know for sure, but If these highly asymmetric galaxies are mergers, than this merger fraction is significantly higher than the merger fraction found at the same redshifts, but for a higher stellar mass sample, giving $f_{m} \sim 0.2$, and measured at the same rest-frame wavelengths \citep[e.g.,][]{ConseliceArnold2009, Duncan2019}. Merger fractions for galaxies at similar masses and redshifts for typical non-lensed field galaxies do not yet exist, so we cannot make a direct comparison yet.  Furthermore, we are probing the rest-frame UV for many of these systems, where galaxies have a more distorted structure, although the effects of this at high redshift are not as pronounced as at lower redshifts \citep[e.g.,][]{Papovich2005}.   These higher asymmetries are however an indication that either mergers or intense star formation are more common for our line emitter systems than a general field sample selection at these redshifts. 

\section{Conclusions}
\label{sec:conclusion}

In this paper we investigate the escape fraction of line emitting galaxies detected in the Hubble Frontier Fields using SHARDS medium-band imaging with the GTC.  We use medium-band filters to find these emission lines sources and ultraviolet HST imaging to measure or place limits on rest-frame Lyman-continuum light. Overall, we discovered a set of 42 new line emitters at $z \sim 2-3.5$ in our SHARDS data that have corresponding coverage of rest-frame UV imaging from the Hubble Space Telescope at redshifts $z = 2 - 3.5$.  Our major findings are:

I.  We find that 2 of our 42 systems have a measurable (3.2$\sigma$) escape fraction. These galaxies are relatively low mass systems (M$_* < 10^9$ M$_{\odot}$), but not at the extreme low-mass end of our sample.  As such our results provide tentative evidence supporting the idea that low-mass galaxies might be an important reionization source.  However, we would expect to have a higher Lyman-continuum fraction, or more sources detected, if lower-mass line emitting galaxies are indeed responsible for producing the energy driving reionization and if our sample are analogs of these \citep[e.g.,][]{Duncan2015}.

II. We stack the 40 galaxies in our sample which do not have $>$2$\sigma$ detections into bins based on cluster membership and the HST/UVIS band. We do not find leaking Lyman-continuum in any of the stacks and calculate upper limits using the same methods we use for the individual galaxies. 

III. The observed distribution of $f_{esc}^{rel}$ with galaxy mass and UV magnitude can be accounted for through a combination of Lyman-continuum detection threshold limits, and uncertainties in the intrinsic flux ratio at lower masses.  We are unable to find or rule out any trends with these properties.

IV. We find that almost half of our candidate emission line galaxies show signs of distorted structures based on an asymmetry index cut.  This cut allows us to investigate the nature of galaxies with distorted light structures which can result from mergers or intense star formation.  However, when we stack these systems we find no statistical difference in escape fraction limits for the asymmetric vs. non-asymmetric galaxies.

Ultimately, our goal is to examine more distant low-mass galaxies at $z > 7$ to determine if they have properties similar to the ones we study in this paper. With JWST we will study the analogs of these systems in the epoch of reionization.  Further examination of similar low mass systems at similar redshifts will also provide more evidence for or against the idea that low mass galaxies, and possibly those involved in mergers or other dynamical events, are those responsible for producing the flux needed for reionization.

\section*{Acknowledgements}

 We thank the referee for their comments which significantly improved the presentation of this paper. This work was supported by the Science and Technology Facilities Council in the form of a studentship to AG.  We also acknowledge support from the ERC Advanced Investigator Grant EPOCHS (788113), as well as a studentship from STFC. LF acknowledges funding from Coordena\c{c}\~{a}o de Aperfei\c{c}oamento de Pessoal de N\'{i}vel Superior - Brasil (CAPES) - Finance Code 001. DC is a Ramon-Cajal Researcher and is supported by the Ministerio de Ciencia, Innovaci\'{o}n y Universidades (MICIU/FEDER) under research grant PGC2018-094975-C21. LR-M acknowledges the support from grant PRIN MIUR2017-20173ML3WW\_001 and DRG thanks CONACyT for the research grant CB-A1-S-22784. This paper is (partly) based on SHARDS-FF data. SHARDS-FF is currently funded by Spanish Government grant PGC2018-093499-BI00. Based on observations made with the Gran Telescopio Canarias (GTC), installed at the Spanish Observatorio del Roque de los Muchachos of the Instituto de Astrof\'{i}sica de Canarias, in the island of La Palma.  This work is also based partially on data and catalog products from HFF-DeepSpace, funded by the National Science Foundation and Space Telescope Science Institute (operated by the Association of Universities for Research in Astronomy, Inc., under NASA contract NAS5-26555).

\software{galclean (\citet{Ferreira2018}), astropy (\citet{Astropy}), FAST (\citep{Kriek2009}), Scipy+Numpy (\citet{scipy2020}), EAZY (\citet{Brammer2008}), OSIRIS pipeline (\citep{Perez2013})}

\bibliography{reference}{}

\begin{thebibliography}{}
\expandafter\ifx\csname natexlab\endcsname\relax\def\natexlab#1{#1}\fi
\providecommand{\url}[1]{\href{#1}{#1}}
\providecommand{\dodoi}[1]{doi:~\href{http://doi.org/#1}{\nolinkurl{#1}}}
\providecommand{\doeprint}[1]{\href{http://ascl.net/#1}{\nolinkurl{http://ascl.net/#1}}}
\providecommand{\doarXiv}[1]{\href{https://arxiv.org/abs/#1}{\nolinkurl{https://arxiv.org/abs/#1}}}

\bibitem[{{Adams} {et~al.}(2022){Adams}, {Conselice}, {Ferreira}, {Austin},
  {Trussler}, {Juod{\v{z}}balis}, {Wilkins}, {Caruana}, {Dayal}, {Verma}, \&
  {Vijayan}}]{Adams2022}
{Adams}, N.~J., {Conselice}, C.~J., {Ferreira}, L., {et~al.} 2022, arXiv
  e-prints, arXiv:2207.11217.
\newblock \doarXiv{2207.11217}

\bibitem[{{Anderson} {et~al.}(2017){Anderson}, {Governato}, {Karcher}, {Quinn},
  \& {Wadsley}}]{Anderson2017}
{Anderson}, L., {Governato}, F., {Karcher}, M., {Quinn}, T., \& {Wadsley}, J.
  2017, \mnras, 468, 4077, \dodoi{10.1093/mnras/stx709}

\bibitem[{{Astropy Collaboration} {et~al.}(2013){Astropy Collaboration},
  {Robitaille}, {Tollerud}, {Greenfield}, {Droettboom}, {Bray}, {Aldcroft},
  {Davis}, {Ginsburg}, {Price-Whelan}, {Kerzendorf}, {Conley}, {Crighton},
  {Barbary}, {Muna}, {Ferguson}, {Grollier}, {Parikh}, {Nair}, {Unther},
  {Deil}, {Woillez}, {Conseil}, {Kramer}, {Turner}, {Singer}, {Fox}, {Weaver},
  {Zabalza}, {Edwards}, {Azalee Bostroem}, {Burke}, {Casey}, {Crawford},
  {Dencheva}, {Ely}, {Jenness}, {Labrie}, {Lim}, {Pierfederici}, {Pontzen},
  {Ptak}, {Refsdal}, {Servillat}, \& {Streicher}}]{Astropy}
{Astropy Collaboration}, {Robitaille}, T.~P., {Tollerud}, E.~J., {et~al.} 2013,
  \aap, 558, A33, \dodoi{10.1051/0004-6361/201322068}

\bibitem[{{Atek} {et~al.}(2022){Atek}, {Furtak}, {Oesch}, {van Dokkum},
  {Reddy}, {Contini}, {Illingworth}, \& {Wilkins}}]{atek2022}
{Atek}, H., {Furtak}, L.~J., {Oesch}, P., {et~al.} 2022, \mnras, 511, 4464,
  \dodoi{10.1093/mnras/stac360}

\bibitem[{{Bershady} {et~al.}(2000){Bershady}, {Jangren}, \&
  {Conselice}}]{Bershady2000}
{Bershady}, M.~A., {Jangren}, A., \& {Conselice}, C.~J. 2000, \aj, 119, 2645,
  \dodoi{10.1086/301386}

\bibitem[{{Bertin} \& {Arnouts}(1996)}]{Bertin1996}
{Bertin}, E., \& {Arnouts}, S. 1996, \aaps, 117, 393,
  \dodoi{10.1051/aas:1996164}

\bibitem[{{Bhatawdekar} {et~al.}(2019){Bhatawdekar}, {Conselice},
  {Margalef-Bentabol}, \& {Duncan}}]{Bhatawdekar2019}
{Bhatawdekar}, R., {Conselice}, C.~J., {Margalef-Bentabol}, B., \& {Duncan}, K.
  2019, \mnras, 486, 3805, \dodoi{10.1093/mnras/stz866}

\bibitem[{{Bian} {et~al.}(2017){Bian}, {Fan}, {McGreer}, {Cai}, \&
  {Jiang}}]{Bian2017}
{Bian}, F., {Fan}, X., {McGreer}, I., {Cai}, Z., \& {Jiang}, L. 2017, \apjl,
  837, L12, \dodoi{10.3847/2041-8213/aa5ff7}

\bibitem[{{Bluck} {et~al.}(2012){Bluck}, {Conselice}, {Buitrago},
  {Gr{\"u}tzbauch}, {Hoyos}, {Mortlock}, \& {Bauer}}]{bluck2012}
{Bluck}, A. F.~L., {Conselice}, C.~J., {Buitrago}, F., {et~al.} 2012, \apj,
  747, 34, \dodoi{10.1088/0004-637X/747/1/34}

\bibitem[{{Boutsia} {et~al.}(2011){Boutsia}, {Grazian}, {Giallongo}, {Fontana},
  {Pentericci}, {Castellano}, {Zamorani}, {Mignoli}, {Vanzella}, {Fiore},
  {Lilly}, {Gallozzi}, {Testa}, {Paris}, \& {Santini}}]{Boutsia2011}
{Boutsia}, K., {Grazian}, A., {Giallongo}, E., {et~al.} 2011, \apj, 736, 41,
  \dodoi{10.1088/0004-637X/736/1/41}

\bibitem[{{Bouwens} {et~al.}(2016){Bouwens}, {Smit}, {Labb{\'e}}, {Franx},
  {Caruana}, {Oesch}, {Stefanon}, \& {Rasappu}}]{Bouwens2016}
{Bouwens}, R.~J., {Smit}, R., {Labb{\'e}}, I., {et~al.} 2016, \apj, 831, 176,
  \dodoi{10.3847/0004-637X/831/2/176}

\bibitem[{{Bradley} {et~al.}(2022){Bradley}, {Coe}, {Brammer}, {Furtak},
  {Larson}, {Andrade-Santos}, {Bhatawdekar}, {Bradac}, {Broadhurst}, {Carnall},
  {Conselice}, {Diego}, {Frye}, {Fujimoto}, {Y. -Y Hsiao}, {Hutchison}, {Jung},
  {Mahler}, {McCandliss}, {Oguri}, {Postman}, {Sharon}, {Trenti}, {Vanzella},
  {Welch}, {Windhorst}, \& {Zitrin}}]{Bradley2022}
{Bradley}, L.~D., {Coe}, D., {Brammer}, G., {et~al.} 2022, arXiv e-prints,
  arXiv:2210.01777.
\newblock \doarXiv{2210.01777}

\bibitem[{{Brammer} {et~al.}(2008){Brammer}, {van Dokkum}, \&
  {Coppi}}]{Brammer2008}
{Brammer}, G.~B., {van Dokkum}, P.~G., \& {Coppi}, P. 2008, \apj, 686, 1503,
  \dodoi{10.1086/591786}

\bibitem[{{Bridge} {et~al.}(2010){Bridge}, {Teplitz}, {Siana}, {Scarlata},
  {Conselice}, {Ferguson}, {Brown}, {Salvato}, {Rudie}, {de Mello}, {Colbert},
  {Gardner}, {Giavalisco}, \& {Armus}}]{Bridge2010}
{Bridge}, C.~R., {Teplitz}, H.~I., {Siana}, B., {et~al.} 2010, \apj, 720, 465,
  \dodoi{10.1088/0004-637X/720/1/465}

\bibitem[{{Bunker} {et~al.}(1995){Bunker}, {Warren}, {Hewett}, \&
  {Clements}}]{Bunker1995}
{Bunker}, A.~J., {Warren}, S.~J., {Hewett}, P.~C., \& {Clements}, D.~L. 1995,
  \mnras, 273, 513, \dodoi{10.1093/mnras/273.2.513}

\bibitem[{{Ceverino} {et~al.}(2019){Ceverino}, {Klessen}, \&
  {Glover}}]{Ceverino2019}
{Ceverino}, D., {Klessen}, R.~S., \& {Glover}, S. C.~O. 2019, \mnras, 484,
  1366, \dodoi{10.1093/mnras/stz079}

\bibitem[{{Chen} {et~al.}(2014){Chen}, {Wise}, {Norman}, {Xu}, \&
  {O'Shea}}]{Chen2014}
{Chen}, P., {Wise}, J.~H., {Norman}, M.~L., {Xu}, H., \& {O'Shea}, B.~W. 2014,
  \apj, 795, 144, \dodoi{10.1088/0004-637X/795/2/144}

\bibitem[{{Conselice}(2003)}]{Conselice2003}
{Conselice}, C.~J. 2003, \apjs, 147, 1, \dodoi{10.1086/375001}

\bibitem[{Conselice(2014)}]{Conselice2014}
Conselice, C.~J. 2014, Annual Review of Astronomy and Astrophysics, 52, 291,
  \dodoi{10.1146/annurev-astro-081913-040037}

\bibitem[{{Conselice} \& {Arnold}(2009)}]{ConseliceArnold2009}
{Conselice}, C.~J., \& {Arnold}, J. 2009, \mnras, 397, 208,
  \dodoi{10.1111/j.1365-2966.2009.14959.x}

\bibitem[{Conselice {et~al.}(2000)Conselice, Bershady, \&
  Jangren}]{Conselice2000}
Conselice, C.~J., Bershady, M.~A., \& Jangren, A. 2000, \apj, 529, 886,
  \dodoi{10.1086/308300}

\bibitem[{{Conselice} {et~al.}(2003){Conselice}, {Chapman}, \&
  {Windhorst}}]{Conselice2003a}
{Conselice}, C.~J., {Chapman}, S.~C., \& {Windhorst}, R.~A. 2003, \apjl, 596,
  L5, \dodoi{10.1086/379109}

\bibitem[{{Conselice} {et~al.}(2008){Conselice}, {Rajgor}, \&
  {Myers}}]{Conselice2008}
{Conselice}, C.~J., {Rajgor}, S., \& {Myers}, R. 2008, \mnras, 386, 909,
  \dodoi{10.1111/j.1365-2966.2008.13069.x}

\bibitem[{{Conselice} {et~al.}(2016){Conselice}, {Wilkinson}, {Duncan}, \&
  {Mortlock}}]{conselice2016}
{Conselice}, C.~J., {Wilkinson}, A., {Duncan}, K., \& {Mortlock}, A. 2016,
  \apj, 830, 83, \dodoi{10.3847/0004-637X/830/2/83}

\bibitem[{{de Barros} {et~al.}(2016){de Barros}, {Vanzella}, {Amor{\'\i}n},
  {Castellano}, {Siana}, {Grazian}, {Suh}, {Balestra}, {Vignali}, {Verhamme},
  {Zamorani}, {Mignoli}, {Hasinger}, {Comastri}, {Pentericci},
  {P{\'e}rez-Montero}, {Fontana}, {Giavalisco}, \& {Gilli}}]{deBarros2016}
{de Barros}, S., {Vanzella}, E., {Amor{\'\i}n}, R., {et~al.} 2016, \aap, 585,
  A51, \dodoi{10.1051/0004-6361/201527046}

\bibitem[{{Dopita} {et~al.}(2011){Dopita}, {Krauss}, {Sutherland },
  {Kobayashi}, \& {Lineweaver}}]{Dopita2011}
{Dopita}, M.~A., {Krauss}, L.~M., {Sutherland }, R.~S., {Kobayashi}, C., \&
  {Lineweaver}, C.~H. 2011, \apss, 335, 345, \dodoi{10.1007/s10509-011-0786-7}

\bibitem[{{Duncan} \& {Conselice}(2015)}]{Duncan2015}
{Duncan}, K., \& {Conselice}, C.~J. 2015, \mnras, 451, 2030,
  \dodoi{10.1093/mnras/stv1049}

\bibitem[{{Duncan} {et~al.}(2019){Duncan}, {Conselice}, {Mundy}, {Bell},
  {Donley}, {Galametz}, {Guo}, {Grogin}, {Hathi}, {Kartaltepe}, {Kocevski},
  {Koekemoer}, {P{\'e}rez-Gonz{\'a}lez}, {Mantha}, {Snyder}, \&
  {Stefanon}}]{Duncan2019}
{Duncan}, K., {Conselice}, C.~J., {Mundy}, C., {et~al.} 2019, \apj, 876, 110,
  \dodoi{10.3847/1538-4357/ab148a}

\bibitem[{{Fan} {et~al.}(2006){Fan}, {Strauss}, {Becker}, {White}, {Gunn},
  {Knapp}, {Richards}, {Schneider}, {Brinkmann}, \& {Fukugita}}]{Fan2006}
{Fan}, X., {Strauss}, M.~A., {Becker}, R.~H., {et~al.} 2006, \aj, 132, 117,
  \dodoi{10.1086/504836}

\bibitem[{{Ferrarese} {et~al.}(2012){Ferrarese}, {C{\^o}t{\'e}}, {Cuilland re},
  {Gwyn}, {Peng}, {MacArthur}, {Duc}, {Boselli}, {Mei}, {Erben}, {McConnachie},
  {Durrell}, {Mihos}, {Jord{\'a}n}, {Lan{\c{c}}on}, {Puzia}, {Emsellem},
  {Balogh}, {Blakeslee}, {van Waerbeke}, {Gavazzi}, {Vollmer}, {Kavelaars},
  {Woods}, {Ball}, {Boissier}, {Courteau}, {Ferriere}, {Gavazzi},
  {Hildebrandt}, {Hudelot}, {Huertas-Company}, {Liu}, {McLaughlin}, {Mellier},
  {Milkeraitis}, {Schade}, {Balkowski}, {Bournaud}, {Carlberg}, {Chapman},
  {Hoekstra}, {Peng}, {Sawicki}, {Simard}, {Taylor}, {Tully}, {van Driel},
  {Wilson}, {Burdullis}, {Mahoney}, \& {Manset}}]{Ferrarese2012}
{Ferrarese}, L., {C{\^o}t{\'e}}, P., {Cuilland re}, J.-C., {et~al.} 2012,
  \apjs, 200, 4, \dodoi{10.1088/0067-0049/200/1/4}

\bibitem[{{Ferrari} {et~al.}(2015){Ferrari}, {de Carvalho}, \&
  {Trevisan}}]{Morfometryka}
{Ferrari}, F., {de Carvalho}, R.~R., \& {Trevisan}, M. 2015, \apj, 814, 55,
  \dodoi{10.1088/0004-637X/814/1/55}

\bibitem[{{Ferreira} {et~al.}(2018){Ferreira}, {Ferrari}, {Griffiths}, \&
  {Tohill}}]{Ferreira2018}
{Ferreira}, L., {Ferrari}, F., {Griffiths}, A., \& {Tohill}, C.-B. 2018,
  {galclean: v1.0.0}, v1.0.0, Zenodo,  Zenodo, \dodoi{10.5281/zenodo.4004571}

\bibitem[{{Finkelstein} {et~al.}(2012){Finkelstein}, {Papovich}, {Ryan},
  {Pawlik}, {Dickinson}, {Ferguson}, {Finlator}, {Koekemoer}, {Giavalisco},
  {Cooray}, {Dunlop}, {Faber}, {Grogin}, {Kocevski}, \&
  {Newman}}]{Finkelstein2012}
{Finkelstein}, S.~L., {Papovich}, C., {Ryan}, R.~E., {et~al.} 2012, \apj, 758,
  93, \dodoi{10.1088/0004-637X/758/2/93}

\bibitem[{{Finkelstein} {et~al.}(2019){Finkelstein}, {D'Aloisio},
  {Paardekooper}, {Ryan}, {Behroozi}, {Finlator}, {Livermore}, {Upton
  Sanderbeck}, {Dalla Vecchia}, \& {Khochfar}}]{Finkelstein2019}
{Finkelstein}, S.~L., {D'Aloisio}, A., {Paardekooper}, J.-P., {et~al.} 2019,
  \apj, 879, 36, \dodoi{10.3847/1538-4357/ab1ea8}

\bibitem[{{Fletcher} {et~al.}(2019){Fletcher}, {Tang}, {Robertson}, {Nakajima},
  {Ellis}, {Stark}, \& {Inoue}}]{fletcher2019}
{Fletcher}, T.~J., {Tang}, M., {Robertson}, B.~E., {et~al.} 2019, \apj, 878,
  87, \dodoi{10.3847/1538-4357/ab2045}

\bibitem[{{Flury} {et~al.}(2022){Flury}, {Jaskot}, {Ferguson}, {Worseck},
  {Makan}, {Chisholm}, {Saldana-Lopez}, {Schaerer}, {McCandliss}, {Wang},
  {Ford}, {Heckman}, {Ji}, {Giavalisco}, {Amorin}, {Atek}, {Blaizot},
  {Borthakur}, {Carr}, {Castellano}, {Cristiani}, {De Barros}, {Dickinson},
  {Finkelstein}, {Fleming}, {Fontanot}, {Garel}, {Grazian}, {Hayes}, {Henry},
  {Mauerhofer}, {Micheva}, {Oey}, {Ostlin}, {Papovich}, {Pentericci},
  {Ravindranath}, {Rosdahl}, {Rutkowski}, {Santini}, {Scarlata}, {Teplitz},
  {Thuan}, {Trebitsch}, {Vanzella}, {Verhamme}, \& {Xu}}]{flury2022}
{Flury}, S.~R., {Jaskot}, A.~E., {Ferguson}, H.~C., {et~al.} 2022, \apjs, 260,
  1, \dodoi{10.3847/1538-4365/ac5331}

\bibitem[{{Genzel} {et~al.}(2014){Genzel}, {F{\"o}rster Schreiber}, {Rosario},
  {Lang}, {Lutz}, {Wisnioski}, {Wuyts}, {Wuyts}, {Bandara}, {Bender}, {Berta},
  {Kurk}, {Mendel}, {Tacconi}, {Wilman}, {Beifiori}, {Brammer}, {Burkert},
  {Buschkamp}, {Chan}, {Carollo}, {Davies}, {Eisenhauer}, {Fabricius},
  {Fossati}, {Kriek}, {Kulkarni}, {Lilly}, {Mancini}, {Momcheva}, {Naab},
  {Nelson}, {Renzini}, {Saglia}, {Sharples}, {Sternberg}, {Tacchella}, \& {van
  Dokkum}}]{Genzel2014}
{Genzel}, R., {F{\"o}rster Schreiber}, N.~M., {Rosario}, D., {et~al.} 2014,
  \apj, 796, 7, \dodoi{10.1088/0004-637X/796/1/7}

\bibitem[{{Giallongo} {et~al.}(2015){Giallongo}, {Grazian}, {Fiore}, {Fontana},
  {Pentericci}, {Vanzella}, {Dickinson}, {Kocevski}, {Castellano}, {Cristiani},
  {Ferguson}, {Finkelstein}, {Grogin}, {Hathi}, {Koekemoer}, {Newman}, \&
  {Salvato}}]{Giallongo2015}
{Giallongo}, E., {Grazian}, A., {Fiore}, F., {et~al.} 2015, \aap, 578, A83,
  \dodoi{10.1051/0004-6361/201425334}

\bibitem[{{Glikman} {et~al.}(2011){Glikman}, {Djorgovski}, {Stern}, {Dey},
  {Jannuzi}, \& {Lee}}]{Glikman2011}
{Glikman}, E., {Djorgovski}, S.~G., {Stern}, D., {et~al.} 2011, \apjl, 728,
  L26, \dodoi{10.1088/2041-8205/728/2/L26}

\bibitem[{{Grazian} {et~al.}(2012){Grazian}, {Castellano}, {Fontana},
  {Pentericci}, {Dunlop}, {McLure}, {Koekemoer}, {Dickinson}, {Faber},
  {Ferguson}, {Galametz}, {Giavalisco}, {Grogin}, {Hathi}, {Kocevski}, {Lai},
  {Newman}, \& {Vanzella}}]{Grazian2012}
{Grazian}, A., {Castellano}, M., {Fontana}, A., {et~al.} 2012, \aap, 547, A51,
  \dodoi{10.1051/0004-6361/201219669}

\bibitem[{{Grazian} {et~al.}(2017){Grazian}, {Giallongo}, {Paris}, {Boutsia},
  {Dickinson}, {Santini}, {Windhorst}, {Jansen}, {Cohen}, {Ashcraft},
  {Scarlata}, {Rutkowski}, {Vanzella}, {Cusano}, {Cristiani}, {Giavalisco},
  {Ferguson}, {Koekemoer}, {Grogin}, {Castellano}, {Fiore}, {Fontana},
  {Marchi}, {Pedichini}, {Pentericci}, {Amor{\'\i}n}, {Barro}, {Bonchi},
  {Bongiorno}, {Faber}, {Fumana}, {Galametz}, {Guaita}, {Kocevski}, {Merlin},
  {Nonino}, {O'Connell}, {Pilo}, {Ryan}, {Sani}, {Speziali}, {Testa}, {Weiner},
  \& {Yan}}]{Grazian2017}
{Grazian}, A., {Giallongo}, E., {Paris}, D., {et~al.} 2017, \aap, 602, A18,
  \dodoi{10.1051/0004-6361/201730447}

\bibitem[{{Grazian} {et~al.}(2018){Grazian}, {Giallongo}, {Boutsia},
  {Cristiani}, {Vanzella}, {Scarlata}, {Santini}, {Pentericci}, {Merlin},
  {Menci}, {Fontanot}, {Fontana}, {Fiore}, {Civano}, {Castellano}, {Brusa},
  {Bonchi}, {Carini}, {Cusano}, {Faccini}, {Garilli}, {Marchetti}, {Rossi}, \&
  {Speziali}}]{Grazian2018}
{Grazian}, A., {Giallongo}, E., {Boutsia}, K., {et~al.} 2018, \aap, 613, A44,
  \dodoi{10.1051/0004-6361/201732385}

\bibitem[{{Griffiths} {et~al.}(2021){Griffiths}, {Conselice}, {Ferreira},
  {Ceverino}, {Rosa-Gonz{\'a}lez}, {Huertas-Company}, {Pampliega},
  {P{\'e}rez-Gonz{\'a}lez}, {Sanchez}, \& {Vega}}]{Griffiths2021}
{Griffiths}, A., {Conselice}, C.~J., {Ferreira}, L., {et~al.} 2021, \mnras,
  508, 3860, \dodoi{10.1093/mnras/stab2566}

\bibitem[{{Grogin} {et~al.}(2011){Grogin}, {Kocevski}, {Faber}, {Ferguson},
  {Koekemoer}, {Riess}, {Acquaviva}, {Alexander}, {Almaini}, {Ashby}, {Barden},
  {Bell}, {Bournaud}, {Brown}, {Caputi}, {Casertano}, {Cassata}, {Castellano},
  {Challis}, {Chary}, {Cheung}, {Cirasuolo}, {Conselice}, {Roshan Cooray},
  {Croton}, {Daddi}, {Dahlen}, {Dav{\'e}}, {de Mello}, {Dekel}, {Dickinson},
  {Dolch}, {Donley}, {Dunlop}, {Dutton}, {Elbaz}, {Fazio}, {Filippenko},
  {Finkelstein}, {Fontana}, {Gardner}, {Garnavich}, {Gawiser}, {Giavalisco},
  {Grazian}, {Guo}, {Hathi}, {H{\"a}ussler}, {Hopkins}, {Huang}, {Huang},
  {Jha}, {Kartaltepe}, {Kirshner}, {Koo}, {Lai}, {Lee}, {Li}, {Lotz}, {Lucas},
  {Madau}, {McCarthy}, {McGrath}, {McIntosh}, {McLure}, {Mobasher},
  {Moustakas}, {Mozena}, {Nandra}, {Newman}, {Niemi}, {Noeske}, {Papovich},
  {Pentericci}, {Pope}, {Primack}, {Rajan}, {Ravindranath}, {Reddy}, {Renzini},
  {Rix}, {Robaina}, {Rodney}, {Rosario}, {Rosati}, {Salimbeni}, {Scarlata},
  {Siana}, {Simard}, {Smidt}, {Somerville}, {Spinrad}, {Straughn}, {Strolger},
  {Telford}, {Teplitz}, {Trump}, {van der Wel}, {Villforth}, {Wechsler},
  {Weiner}, {Wiklind}, {Wild}, {Wilson}, {Wuyts}, {Yan}, \& {Yun}}]{Grogin2011}
{Grogin}, N.~A., {Kocevski}, D.~D., {Faber}, S.~M., {et~al.} 2011, \apjs, 197,
  35, \dodoi{10.1088/0067-0049/197/2/35}

\bibitem[{{Hassan} {et~al.}(2016){Hassan}, {Dav{\'e}}, {Finlator}, \&
  {Santos}}]{Hassan2016}
{Hassan}, S., {Dav{\'e}}, R., {Finlator}, K., \& {Santos}, M.~G. 2016, \mnras,
  457, 1550, \dodoi{10.1093/mnras/stv3001}

\bibitem[{{Inoue} {et~al.}(2005){Inoue}, {Iwata}, {Deharveng}, {Buat}, \&
  {Burgarella}}]{Inoue2005}
{Inoue}, A.~K., {Iwata}, I., {Deharveng}, J.~M., {Buat}, V., \& {Burgarella},
  D. 2005, \aap, 435, 471, \dodoi{10.1051/0004-6361:20041769}

\bibitem[{{Inoue} {et~al.}(2014){Inoue}, {Shimizu}, {Iwata}, \&
  {Tanaka}}]{Inoue2014}
{Inoue}, A.~K., {Shimizu}, I., {Iwata}, I., \& {Tanaka}, M. 2014, \mnras, 442,
  1805, \dodoi{10.1093/mnras/stu936}

\bibitem[{{Iwata} {et~al.}(2009){Iwata}, {Inoue}, {Matsuda}, {Furusawa},
  {Hayashino}, {Kousai}, {Akiyama}, {Yamada}, {Burgarella}, \&
  {Deharveng}}]{Iwata2009}
{Iwata}, I., {Inoue}, A.~K., {Matsuda}, Y., {et~al.} 2009, \apj, 692, 1287,
  \dodoi{10.1088/0004-637X/692/2/1287}

\bibitem[{{Izotov} {et~al.}(2016){Izotov}, {Orlitov{\'a}}, {Schaerer}, {Thuan},
  {Verhamme}, {Guseva}, \& {Worseck}}]{Izotov2016}
{Izotov}, Y.~I., {Orlitov{\'a}}, I., {Schaerer}, D., {et~al.} 2016, \nat, 529,
  178, \dodoi{10.1038/nature16456}

\bibitem[{{Ji} {et~al.}(2020){Ji}, {Giavalisco}, {Vanzella}, {Siana},
  {Pentericci}, {Jaskot}, {Liu}, {Nonino}, {Ferguson}, {Castellano},
  {Mannucci}, {Schaerer}, {Fynbo}, {Papovich}, {Carnall}, {Amorin}, {Simons},
  {Hathi}, {Cullen}, \& {McLeod}}]{Ji2020}
{Ji}, Z., {Giavalisco}, M., {Vanzella}, E., {et~al.} 2020, \apj, 888, 109,
  \dodoi{10.3847/1538-4357/ab5fdc}

\bibitem[{{Jia} {et~al.}(2011){Jia}, {Ptak}, {Heckman}, {Overzier},
  {Hornschemeier}, \& {LaMassa}}]{Jia2011}
{Jia}, J., {Ptak}, A., {Heckman}, T.~M., {et~al.} 2011, \apj, 731, 55,
  \dodoi{10.1088/0004-637X/731/1/55}

\bibitem[{{Kaaret} {et~al.}(2017){Kaaret}, {Brorby}, {Casella}, \&
  {Prestwich}}]{Kaaret2017}
{Kaaret}, P., {Brorby}, M., {Casella}, L., \& {Prestwich}, A.~H. 2017, \mnras,
  471, 4234, \dodoi{10.1093/mnras/stx1945}

\bibitem[{{Koekemoer} {et~al.}(2011){Koekemoer}, {Faber}, {Ferguson}, {Grogin},
  {Kocevski}, {Koo}, {Lai}, {Lotz}, {Lucas}, {McGrath}, {Ogaz}, {Rajan},
  {Riess}, {Rodney}, {Strolger}, {Casertano}, {Castellano}, {Dahlen},
  {Dickinson}, {Dolch}, {Fontana}, {Giavalisco}, {Grazian}, {Guo}, {Hathi},
  {Huang}, {van der Wel}, {Yan}, {Acquaviva}, {Alexander}, {Almaini}, {Ashby},
  {Barden}, {Bell}, {Bournaud}, {Brown}, {Caputi}, {Cassata}, {Challis},
  {Chary}, {Cheung}, {Cirasuolo}, {Conselice}, {Roshan Cooray}, {Croton},
  {Daddi}, {Dav{\'e}}, {de Mello}, {de Ravel}, {Dekel}, {Donley}, {Dunlop},
  {Dutton}, {Elbaz}, {Fazio}, {Filippenko}, {Finkelstein}, {Frazer}, {Gardner},
  {Garnavich}, {Gawiser}, {Gruetzbauch}, {Hartley}, {H{\"a}ussler},
  {Herrington}, {Hopkins}, {Huang}, {Jha}, {Johnson}, {Kartaltepe},
  {Khostovan}, {Kirshner}, {Lani}, {Lee}, {Li}, {Madau}, {McCarthy},
  {McIntosh}, {McLure}, {McPartland}, {Mobasher}, {Moreira}, {Mortlock},
  {Moustakas}, {Mozena}, {Nandra}, {Newman}, {Nielsen}, {Niemi}, {Noeske},
  {Papovich}, {Pentericci}, {Pope}, {Primack}, {Ravindranath}, {Reddy},
  {Renzini}, {Rix}, {Robaina}, {Rosario}, {Rosati}, {Salimbeni}, {Scarlata},
  {Siana}, {Simard}, {Smidt}, {Snyder}, {Somerville}, {Spinrad}, {Straughn},
  {Telford}, {Teplitz}, {Trump}, {Vargas}, {Villforth}, {Wagner}, {Wandro},
  {Wechsler}, {Weiner}, {Wiklind}, {Wild}, {Wilson}, {Wuyts}, \&
  {Yun}}]{Koekemoer2011}
{Koekemoer}, A.~M., {Faber}, S.~M., {Ferguson}, H.~C., {et~al.} 2011, \apjs,
  197, 36, \dodoi{10.1088/0067-0049/197/2/36}

\bibitem[{{Kondapally} {et~al.}(2018){Kondapally}, {Russell}, {Conselice}, \&
  {Penny}}]{Kondapally2018}
{Kondapally}, R., {Russell}, G.~A., {Conselice}, C.~J., \& {Penny}, S.~J. 2018,
  \mnras, 481, 1759, \dodoi{10.1093/mnras/sty2333}

\bibitem[{{Kriek} {et~al.}(2009){Kriek}, {van Dokkum}, {Labb{\'e}}, {Franx},
  {Illingworth}, {Marchesini}, \& {Quadri}}]{Kriek2009}
{Kriek}, M., {van Dokkum}, P.~G., {Labb{\'e}}, I., {et~al.} 2009, \apj, 700,
  221, \dodoi{10.1088/0004-637X/700/1/221}

\bibitem[{{Lagattuta} {et~al.}(2019){Lagattuta}, {Richard}, {Bauer},
  {Cl{\'e}ment}, {Mahler}, {Soucail}, {Carton}, {Kneib}, {Laporte}, {Martinez},
  {Patr{\'\i}cio}, {Payne}, {Pell{\'o}}, {Schmidt}, \& {de la
  Vieuville}}]{Lagattuta2019}
{Lagattuta}, D.~J., {Richard}, J., {Bauer}, F.~E., {et~al.} 2019, \mnras, 485,
  3738, \dodoi{10.1093/mnras/stz620}

\bibitem[{{Leisman} {et~al.}(2017){Leisman}, {Haynes}, {Janowiecki},
  {Hallenbeck}, {J{\'o}zsa}, {Giovanelli}, {Adams}, {Bernal Neira}, {Cannon},
  {Janesh}, {Rhode}, \& {Salzer}}]{Leisman2017}
{Leisman}, L., {Haynes}, M.~P., {Janowiecki}, S., {et~al.} 2017, \apj, 842,
  133, \dodoi{10.3847/1538-4357/aa7575}

\bibitem[{{Leitherer} {et~al.}(2016){Leitherer}, {Hernandez}, {Lee}, \&
  {Oey}}]{Leitherer2016}
{Leitherer}, C., {Hernandez}, S., {Lee}, J.~C., \& {Oey}, M.~S. 2016, \apj,
  823, 64, \dodoi{10.3847/0004-637X/823/1/64}

\bibitem[{{Leitherer} {et~al.}(1999){Leitherer}, {Schaerer}, {Goldader},
  {Delgado}, {Robert}, {Kune}, {de Mello}, {Devost}, \&
  {Heckman}}]{Leitherer1999}
{Leitherer}, C., {Schaerer}, D., {Goldader}, J.~D., {et~al.} 1999, \apjs, 123,
  3, \dodoi{10.1086/313233}

\bibitem[{{Lotz} {et~al.}(2017){Lotz}, {Koekemoer}, {Coe}, {Grogin}, {Capak},
  {Mack}, {Anderson}, {Avila}, {Barker}, {Borncamp}, {Brammer}, {Durbin},
  {Gunning}, {Hilbert}, {Jenkner}, {Khandrika}, {Levay}, {Lucas}, {MacKenty},
  {Ogaz}, {Porterfield}, {Reid}, {Robberto}, {Royle}, {Smith},
  {Storrie-Lombardi}, {Sunnquist}, {Surace}, {Taylor}, {Williams}, {Bullock},
  {Dickinson}, {Finkelstein}, {Natarajan}, {Richard}, {Robertson}, {Tumlinson},
  {Zitrin}, {Flanagan}, {Sembach}, {Soifer}, \& {Mountain}}]{Lotz2017}
{Lotz}, J.~M., {Koekemoer}, A., {Coe}, D., {et~al.} 2017, \apj, 837, 97,
  \dodoi{10.3847/1538-4357/837/1/97}

\bibitem[{{Ma} {et~al.}(2016){Ma}, {Hopkins}, {Kasen}, {Quataert},
  {Faucher-Gigu{\`e}re}, {Kere{\v{s}}}, {Murray}, \& {Strom}}]{Ma2016}
{Ma}, X., {Hopkins}, P.~F., {Kasen}, D., {et~al.} 2016, \mnras, 459, 3614,
  \dodoi{10.1093/mnras/stw941}

\bibitem[{{Madau} \& {Haardt}(2015)}]{Madau2015}
{Madau}, P., \& {Haardt}, F. 2015, \apjl, 813, L8,
  \dodoi{10.1088/2041-8205/813/1/L8}

\bibitem[{{Marchi} {et~al.}(2018){Marchi}, {Pentericci}, {Guaita}, {Schaerer},
  {Verhamme}, {Castellano}, {Ribeiro}, {Garilli}, {Le F{\`e}vre}, {Amorin},
  {Bardelli}, {Cassata}, {Durkalec}, {Grazian}, {Hathi}, {Lemaux}, {Maccagni},
  {Vanzella}, \& {Zucca}}]{marchi2018}
{Marchi}, F., {Pentericci}, L., {Guaita}, L., {et~al.} 2018, \aap, 614, A11,
  \dodoi{10.1051/0004-6361/201732133}

\bibitem[{{Marques-Chaves} {et~al.}(2021){Marques-Chaves}, {Schaerer},
  {{\'A}lvarez-M{\'a}rquez}, {Colina}, {Dessauges-Zavadsky},
  {P{\'e}rez-Fournon}, {Saldana-Lopez}, \& {Verhamme}}]{marqueschaves2021}
{Marques-Chaves}, R., {Schaerer}, D., {{\'A}lvarez-M{\'a}rquez}, J., {et~al.}
  2021, \mnras, 507, 524, \dodoi{10.1093/mnras/stab2187}

\bibitem[{{Matthee} {et~al.}(2015){Matthee}, {Sobral}, {Santos},
  {R{\"o}ttgering}, {Darvish}, \& {Mobasher}}]{Matthee2015}
{Matthee}, J., {Sobral}, D., {Santos}, S., {et~al.} 2015, \mnras, 451, 400,
  \dodoi{10.1093/mnras/stv947}

\bibitem[{{Mostardi} {et~al.}(2015){Mostardi}, {Shapley}, {Steidel}, {Trainor},
  {Reddy}, \& {Siana}}]{mostardi2015}
{Mostardi}, R.~E., {Shapley}, A.~E., {Steidel}, C.~C., {et~al.} 2015, \apj,
  810, 107, \dodoi{10.1088/0004-637X/810/2/107}

\bibitem[{{Naidu} {et~al.}(2020){Naidu}, {Tacchella}, {Mason}, {Bose}, {Oesch},
  \& {Conroy}}]{naidu2020}
{Naidu}, R.~P., {Tacchella}, S., {Mason}, C.~A., {et~al.} 2020, \apj, 892, 109,
  \dodoi{10.3847/1538-4357/ab7cc9}

\bibitem[{{Naidu} {et~al.}(2022){Naidu}, {Matthee}, {Oesch}, {Conroy},
  {Sobral}, {Pezzulli}, {Hayes}, {Erb}, {Amor{\'\i}n}, {Gronke}, {Schaerer},
  {Tacchella}, {Kerutt}, {Paulino-Afonso}, {Calhau}, {Llerena}, \&
  {R{\"o}ttgering}}]{Naidu2022}
{Naidu}, R.~P., {Matthee}, J., {Oesch}, P.~A., {et~al.} 2022, \mnras, 510,
  4582, \dodoi{10.1093/mnras/stab3601}

\bibitem[{{Nakajima} {et~al.}(2020){Nakajima}, {Ellis}, {Robertson}, {Tang}, \&
  {Stark}}]{nakajima2020}
{Nakajima}, K., {Ellis}, R.~S., {Robertson}, B.~E., {Tang}, M., \& {Stark},
  D.~P. 2020, \apj, 889, 161, \dodoi{10.3847/1538-4357/ab6604}

\bibitem[{{Nestor} {et~al.}(2011){Nestor}, {Shapley}, {Steidel}, \&
  {Siana}}]{Nestor2011}
{Nestor}, D.~B., {Shapley}, A.~E., {Steidel}, C.~C., \& {Siana}, B. 2011, \apj,
  736, 18, \dodoi{10.1088/0004-637X/736/1/18}

\bibitem[{{Oke}(1974)}]{Oke1974}
{Oke}, J.~B. 1974, \apjs, 27, 21, \dodoi{10.1086/190287}

\bibitem[{{Ouchi} {et~al.}(2008){Ouchi}, {Shimasaku}, {Akiyama}, {Simpson},
  {Saito}, {Ueda}, {Furusawa}, {Sekiguchi}, {Yamada}, {Kodama}, {Kashikawa},
  {Okamura}, {Iye}, {Takata}, {Yoshida}, \& {Yoshida}}]{Ouchi2008}
{Ouchi}, M., {Shimasaku}, K., {Akiyama}, M., {et~al.} 2008, \apjs, 176, 301,
  \dodoi{10.1086/527673}

\bibitem[{{Ouchi} {et~al.}(2010){Ouchi}, {Shimasaku}, {Furusawa}, {Saito},
  {Yoshida}, {Akiyama}, {Ono}, {Yamada}, {Ota}, {Kashikawa}, {Iye}, {Kodama},
  {Okamura}, {Simpson}, \& {Yoshida}}]{Ouchi2010}
{Ouchi}, M., {Shimasaku}, K., {Furusawa}, H., {et~al.} 2010, \apj, 723, 869,
  \dodoi{10.1088/0004-637X/723/1/869}

\bibitem[{{Pahl} {et~al.}(2021){Pahl}, {Shapley}, {Steidel}, {Chen}, \&
  {Reddy}}]{pahl2021}
{Pahl}, A.~J., {Shapley}, A., {Steidel}, C.~C., {Chen}, Y., \& {Reddy}, N.~A.
  2021, arXiv e-prints, arXiv:2104.02081.
\newblock \doarXiv{2104.02081}

\bibitem[{{Papovich} {et~al.}(2005){Papovich}, {Dickinson}, {Giavalisco},
  {Conselice}, \& {Ferguson}}]{Papovich2005}
{Papovich}, C., {Dickinson}, M., {Giavalisco}, M., {Conselice}, C.~J., \&
  {Ferguson}, H.~C. 2005, \apj, 631, 101, \dodoi{10.1086/429120}

\bibitem[{{Parsa} {et~al.}(2018){Parsa}, {Dunlop}, \& {McLure}}]{Parsa2018}
{Parsa}, S., {Dunlop}, J.~S., \& {McLure}, R.~J. 2018, \mnras, 474, 2904,
  \dodoi{10.1093/mnras/stx2887}

\bibitem[{{P{\'e}rez-Gonz{\'a}lez} {et~al.}(2013){P{\'e}rez-Gonz{\'a}lez},
  {Cava}, {Barro}, {Villar}, {Cardiel}, {Ferreras},
  {Rodr{\'{\i}}guez-Espinosa}, {Alonso-Herrero}, {Balcells}, {Cenarro}, {Cepa},
  {Charlot}, {Cimatti}, {Conselice}, {Daddi}, {Donley}, {Elbaz}, {Espino},
  {Gallego}, {Gobat}, {Gonz{\'a}lez-Mart{\'{\i}}n}, {Guzm{\'a}n},
  {Hern{\'a}n-Caballero}, {Mu{\~n}oz-Tu{\~n}{\'o}n}, {Renzini},
  {Rodr{\'{\i}}guez-Zaur{\'{\i}}n}, {Tresse}, {Trujillo}, \&
  {Zamorano}}]{Perez2013}
{P{\'e}rez-Gonz{\'a}lez}, P.~G., {Cava}, A., {Barro}, G., {et~al.} 2013, \apj,
  762, 46, \dodoi{10.1088/0004-637X/762/1/46}

\bibitem[{{Petrosian}(1976)}]{Petrosian}
{Petrosian}, V. 1976, \apjl, 210, L53, \dodoi{10.1086/182301}

\bibitem[{{Planck Collaboration} {et~al.}(2018){Planck Collaboration},
  {Aghanim}, {Akrami}, {Ashdown}, {Aumont}, {Baccigalupi}, {Ballardini},
  {Banday}, {Barreiro}, {Bartolo}, {Basak}, {Battye}, {Benabed}, {Bernard},
  {Bersanelli}, {Bielewicz}, {Bock}, {Bond}, {Borrill}, {Bouchet}, {Boulanger},
  {Bucher}, {Burigana}, {Butler}, {Calabrese}, {Cardoso}, {Carron},
  {Challinor}, {Chiang}, {Chluba}, {Colombo}, {Combet}, {Contreras}, {Crill},
  {Cuttaia}, {de Bernardis}, {de Zotti}, {Delabrouille}, {Delouis}, {Di
  Valentino}, {Diego}, {Dor{\'e}}, {Douspis}, {Ducout}, {Dupac}, {Dusini},
  {Efstathiou}, {Elsner}, {En{\ss}lin}, {Eriksen}, {Fantaye}, {Farhang},
  {Fergusson}, {Fernandez-Cobos}, {Finelli}, {Forastieri}, {Frailis},
  {Franceschi}, {Frolov}, {Galeotta}, {Galli}, {Ganga}, {G{\'e}nova-Santos},
  {Gerbino}, {Ghosh}, {Gonz{\'a}lez-Nuevo}, {G{\'o}rski}, {Gratton},
  {Gruppuso}, {Gudmundsson}, {Hamann}, {Handley}, {Herranz}, {Hivon}, {Huang},
  {Jaffe}, {Jones}, {Karakci}, {Keih{\"a}nen}, {Keskitalo}, {Kiiveri}, {Kim},
  {Kisner}, {Knox}, {Krachmalnicoff}, {Kunz}, {Kurki-Suonio}, {Lagache},
  {Lamarre}, {Lasenby}, {Lattanzi}, {Lawrence}, {Le Jeune}, {Lemos},
  {Lesgourgues}, {Levrier}, {Lewis}, {Liguori}, {Lilje}, {Lilley}, {Lindholm},
  {L{\'o}pez-Caniego}, {Lubin}, {Ma}, {Mac{\'{\i}}as-P{\'e}rez}, {Maggio},
  {Maino}, {Mandolesi}, {Mangilli}, {Marcos-Caballero}, {Maris}, {Martin},
  {Martinelli}, {Mart{\'{\i}}nez-Gonz{\'a}lez}, {Matarrese}, {Mauri}, {McEwen},
  {Meinhold}, {Melchiorri}, {Mennella}, {Migliaccio}, {Millea}, {Mitra},
  {Miville-Desch{\^e}nes}, {Molinari}, {Montier}, {Morgante}, {Moss}, {Natoli},
  {N{\o}rgaard-Nielsen}, {Pagano}, {Paoletti}, {Partridge}, {Patanchon},
  {Peiris}, {Perrotta}, {Pettorino}, {Piacentini}, {Polastri}, {Polenta},
  {Puget}, {Rachen}, {Reinecke}, {Remazeilles}, {Renzi}, {Rocha}, {Rosset},
  {Roudier}, {Rubi{\~n}o-Mart{\'{\i}}n}, {Ruiz-Granados}, {Salvati}, {Sandri},
  {Savelainen}, {Scott}, {Shellard}, {Sirignano}, {Sirri}, {Spencer},
  {Sunyaev}, {Suur-Uski}, {Tauber}, {Tavagnacco}, {Tenti}, {Toffolatti},
  {Tomasi}, {Trombetti}, {Valenziano}, {Valiviita}, {Van Tent}, {Vibert},
  {Vielva}, {Villa}, {Vittorio}, {Wandelt}, {Wehus}, {White}, {White},
  {Zacchei}, \& {Zonca}}]{Planck2018}
{Planck Collaboration}, {Aghanim}, N., {Akrami}, Y., {et~al.} 2018, ArXiv
  e-prints.
\newblock \doarXiv{1807.06209}

\bibitem[{{Prestwich} {et~al.}(2015){Prestwich}, {Jackson}, {Kaaret}, {Brorby},
  {Roberts}, {Saar}, \& {Yukita}}]{Prestwich2015}
{Prestwich}, A.~H., {Jackson}, F., {Kaaret}, P., {et~al.} 2015, \apj, 812, 166,
  \dodoi{10.1088/0004-637X/812/2/166}

\bibitem[{{Puschnig} {et~al.}(2017){Puschnig}, {Hayes}, {{\"O}stlin},
  {Rivera-Thorsen}, {Melinder}, {Cannon}, {Menacho}, {Zackrisson}, {Bergvall},
  \& {Leitet}}]{Puschnig2017}
{Puschnig}, J., {Hayes}, M., {{\"O}stlin}, G., {et~al.} 2017, \mnras, 469,
  3252, \dodoi{10.1093/mnras/stx951}

\bibitem[{{Rivera-Thorsen} {et~al.}(2019){Rivera-Thorsen}, {Dahle}, {Chisholm},
  {Florian}, {Gronke}, {Rigby}, {Gladders}, {Mahler}, {Sharon}, \&
  {Bayliss}}]{RiveraThorsen2019}
{Rivera-Thorsen}, T.~E., {Dahle}, H., {Chisholm}, J., {et~al.} 2019, Science,
  366, 738, \dodoi{10.1126/science.aaw0978}

\bibitem[{{Robertson} {et~al.}(2015){Robertson}, {Ellis}, {Furlanetto}, \&
  {Dunlop}}]{Robertson2015}
{Robertson}, B.~E., {Ellis}, R.~S., {Furlanetto}, S.~R., \& {Dunlop}, J.~S.
  2015, \apjl, 802, L19, \dodoi{10.1088/2041-8205/802/2/L19}

\bibitem[{{Robertson} {et~al.}(2013){Robertson}, {Furlanetto}, {Schneider},
  {Charlot}, {Ellis}, {Stark}, {McLure}, {Dunlop}, {Koekemoer}, {Schenker},
  {Ouchi}, {Ono}, {Curtis-Lake}, {Rogers}, {Bowler}, \&
  {Cirasuolo}}]{Robertson2013}
{Robertson}, B.~E., {Furlanetto}, S.~R., {Schneider}, E., {et~al.} 2013, \apj,
  768, 71, \dodoi{10.1088/0004-637X/768/1/71}

\bibitem[{{Rutkowski} {et~al.}(2016){Rutkowski}, {Scarlata}, {Haardt}, {Siana},
  {Henry}, {Rafelski}, {Hayes}, {Salvato}, {Pahl}, {Mehta}, {Beck}, {Malkan},
  \& {Teplitz}}]{Rutkowski2016}
{Rutkowski}, M.~J., {Scarlata}, C., {Haardt}, F., {et~al.} 2016, \apj, 819, 81,
  \dodoi{10.3847/0004-637X/819/1/81}

\bibitem[{{Rutkowski} {et~al.}(2017){Rutkowski}, {Scarlata}, {Henry}, {Hayes},
  {Mehta}, {Hathi}, {Cohen}, {Windhorst}, {Koekemoer}, {Teplitz}, {Haardt}, \&
  {Siana}}]{Rutkowski2017}
{Rutkowski}, M.~J., {Scarlata}, C., {Henry}, A., {et~al.} 2017, \apjl, 841,
  L27, \dodoi{10.3847/2041-8213/aa733b}

\bibitem[{{Santos} {et~al.}(2016){Santos}, {Sobral}, \& {Matthee}}]{Santos2016}
{Santos}, S., {Sobral}, D., \& {Matthee}, J. 2016, \mnras, 463, 1678,
  \dodoi{10.1093/mnras/stw2076}

\bibitem[{{Saxena} {et~al.}(2022){Saxena}, {Pentericci}, {Ellis}, {Guaita},
  {Calabr{\`o}}, {Schaerer}, {Vanzella}, {Amor{\'\i}n}, {Bolzonella},
  {Castellano}, {Fontanot}, {Hathi}, {Hibon}, {Llerena}, {Mannucci},
  {Saldana-Lopez}, {Talia}, \& {Zamorani}}]{Saxena2022}
{Saxena}, A., {Pentericci}, L., {Ellis}, R.~S., {et~al.} 2022, \mnras, 511,
  120, \dodoi{10.1093/mnras/stab3728}

\bibitem[{{Schaerer} {et~al.}(2022){Schaerer}, {Izotov}, {Worseck}, {Berg},
  {Chisholm}, {Jaskot}, {Nakajima}, {Ravindranath}, {Thuan}, \&
  {Verhamme}}]{Schaerer2022}
{Schaerer}, D., {Izotov}, Y.~I., {Worseck}, G., {et~al.} 2022, \aap, 658, L11,
  \dodoi{10.1051/0004-6361/202243149}

\bibitem[{{Shapley} {et~al.}(2006){Shapley}, {Steidel}, {Pettini},
  {Adelberger}, \& {Erb}}]{Shapley2006}
{Shapley}, A.~E., {Steidel}, C.~C., {Pettini}, M., {Adelberger}, K.~L., \&
  {Erb}, D.~K. 2006, \apj, 651, 688, \dodoi{10.1086/507511}

\bibitem[{{Shapley} {et~al.}(2016){Shapley}, {Steidel}, {Strom},
  {Bogosavljevi{\'c}}, {Reddy}, {Siana}, {Mostardi}, \& {Rudie}}]{Shapley2016}
{Shapley}, A.~E., {Steidel}, C.~C., {Strom}, A.~L., {et~al.} 2016, \apjl, 826,
  L24, \dodoi{10.3847/2041-8205/826/2/L24}

\bibitem[{{Shipley} {et~al.}(2018){Shipley}, {Lange-Vagle}, {Marchesini},
  {Brammer}, {Ferrarese}, {Stefanon}, {Kado-Fong}, {Whitaker}, {Oesch},
  {Feinstein}, {Labb{\'e}}, {Lundgren}, {Martis}, {Muzzin}, {Nedkova},
  {Skelton}, \& {van der Wel}}]{Shipley2018}
{Shipley}, H.~V., {Lange-Vagle}, D., {Marchesini}, D., {et~al.} 2018, \apjs,
  235, 14, \dodoi{10.3847/1538-4365/aaacce}

\bibitem[{{Siana} {et~al.}(2007){Siana}, {Teplitz}, {Colbert}, {Ferguson},
  {Dickinson}, {Brown}, {Conselice}, {de Mello}, {Gardner}, {Giavalisco}, \&
  {Menanteau}}]{Siana2007}
{Siana}, B., {Teplitz}, H.~I., {Colbert}, J., {et~al.} 2007, \apj, 668, 62,
  \dodoi{10.1086/521185}

\bibitem[{{Siana} {et~al.}(2010){Siana}, {Teplitz}, {Ferguson}, {Brown},
  {Giavalisco}, {Dickinson}, {Chary}, {de Mello}, {Conselice}, {Bridge},
  {Gardner}, {Colbert}, \& {Scarlata}}]{Siana2010}
{Siana}, B., {Teplitz}, H.~I., {Ferguson}, H.~C., {et~al.} 2010, \apj, 723,
  241, \dodoi{10.1088/0004-637X/723/1/241}

\bibitem[{{Siana} {et~al.}(2015){Siana}, {Shapley}, {Kulas}, {Nestor},
  {Steidel}, {Teplitz}, {Alavi}, {Brown}, {Conselice}, {Ferguson}, {Dickinson},
  {Giavalisco}, {Colbert}, {Bridge}, {Gardner}, \& {de Mello}}]{Siana2015}
{Siana}, B., {Shapley}, A.~E., {Kulas}, K.~R., {et~al.} 2015, \apj, 804, 17,
  \dodoi{10.1088/0004-637X/804/1/17}

\bibitem[{{Skelton} {et~al.}(2014){Skelton}, {Whitaker}, {Momcheva}, {Brammer},
  {van Dokkum}, {Labb{\'e}}, {Franx}, {van der Wel}, {Bezanson}, {Da Cunha},
  {Fumagalli}, {F{\"o}rster Schreiber}, {Kriek}, {Leja}, {Lundgren}, {Magee},
  {Marchesini}, {Maseda}, {Nelson}, {Oesch}, {Pacifici}, {Patel}, {Price},
  {Rix}, {Tal}, {Wake}, \& {Wuyts}}]{Skelton2014}
{Skelton}, R.~E., {Whitaker}, K.~E., {Momcheva}, I.~G., {et~al.} 2014, \apjs,
  214, 24, \dodoi{10.1088/0067-0049/214/2/24}

\bibitem[{{Smith} {et~al.}(2020){Smith}, {Windhorst}, {Cohen}, {Koekemoer},
  {Jansen}, {White}, {Borthakur}, {Hathi}, {Jiang}, {Rutkowski}, {Ryan},
  {Inoue}, {O'Connell}, {MacKenty}, {Conselice}, \& {Silk}}]{Smith2020}
{Smith}, B.~M., {Windhorst}, R.~A., {Cohen}, S.~H., {et~al.} 2020, arXiv
  e-prints, arXiv:2004.04360.
\newblock \doarXiv{2004.04360}

\bibitem[{{Sobral} {et~al.}(2013){Sobral}, {Smail}, {Best}, {Geach}, {Matsuda},
  {Stott}, {Cirasuolo}, \& {Kurk}}]{Sobral2013}
{Sobral}, D., {Smail}, I., {Best}, P.~N., {et~al.} 2013, \mnras, 428, 1128,
  \dodoi{10.1093/mnras/sts096}

\bibitem[{{Sobral} {et~al.}(2017){Sobral}, {Matthee}, {Best}, {Stroe},
  {R{\"o}ttgering}, {Oteo}, {Smail}, {Morabito}, \&
  {Paulino-Afonso}}]{Sobral2017}
{Sobral}, D., {Matthee}, J., {Best}, P., {et~al.} 2017, \mnras, 466, 1242,
  \dodoi{10.1093/mnras/stw3090}

\bibitem[{{Stark} {et~al.}(2015){Stark}, {Walth}, {Charlot}, {Cl{\'e}ment},
  {Feltre}, {Gutkin}, {Richard}, {Mainali}, {Robertson}, {Siana}, {Tang}, \&
  {Schenker}}]{stark2015}
{Stark}, D.~P., {Walth}, G., {Charlot}, S., {et~al.} 2015, \mnras, 454, 1393,
  \dodoi{10.1093/mnras/stv1907}

\bibitem[{{Steidel} {et~al.}(2018){Steidel}, {Bogosavljevi{\'c}}, {Shapley},
  {Reddy}, {Rudie}, {Pettini}, {Trainor}, \& {Strom}}]{Steidel2018}
{Steidel}, C.~C., {Bogosavljevi{\'c}}, M., {Shapley}, A.~E., {et~al.} 2018,
  \apj, 869, 123, \dodoi{10.3847/1538-4357/aaed28}

\bibitem[{{Steidel} {et~al.}(2001){Steidel}, {Pettini}, \&
  {Adelberger}}]{Steidel2001}
{Steidel}, C.~C., {Pettini}, M., \& {Adelberger}, K.~L. 2001, \apj, 546, 665,
  \dodoi{10.1086/318323}

\bibitem[{{Torres-Alb{\`a}} {et~al.}(2020){Torres-Alb{\`a}}, {Bosch-Ramon}, \&
  {Iwasawa}}]{Torres2020}
{Torres-Alb{\`a}}, N., {Bosch-Ramon}, V., \& {Iwasawa}, K. 2020, \aap, 635,
  A57, \dodoi{10.1051/0004-6361/201936047}

\bibitem[{{Trebitsch} {et~al.}(2017){Trebitsch}, {Blaizot}, {Rosdahl},
  {Devriendt}, \& {Slyz}}]{Trebitsch2017}
{Trebitsch}, M., {Blaizot}, J., {Rosdahl}, J., {Devriendt}, J., \& {Slyz}, A.
  2017, in Galaxy Evolution Across Time, 49, \dodoi{10.5281/zenodo.808185}

\bibitem[{{Trussler} {et~al.}(2022){Trussler}, {Adams}, {Conselice},
  {Ferreira}, {Austin}, {Bhatawdekar}, {Caruana}, {Lovell}, {Roper}, {Verma},
  {Vijayan}, \& {Wilkins}}]{Trussler2022}
{Trussler}, J. A.~A., {Adams}, N.~J., {Conselice}, C.~J., {et~al.} 2022, arXiv
  e-prints, arXiv:2207.14265.
\newblock \doarXiv{2207.14265}

\bibitem[{{Vanzella} {et~al.}(2012){Vanzella}, {Guo}, {Giavalisco}, {Grazian},
  {Castellano}, {Cristiani}, {Dickinson}, {Fontana}, {Nonino}, {Giallongo},
  {Pentericci}, {Galametz}, {Faber}, {Ferguson}, {Grogin}, {Koekemoer},
  {Newman}, \& {Siana}}]{Vanzella2012}
{Vanzella}, E., {Guo}, Y., {Giavalisco}, M., {et~al.} 2012, \apj, 751, 70,
  \dodoi{10.1088/0004-637X/751/1/70}

\bibitem[{{Vanzella} {et~al.}(2016){Vanzella}, {de Barros}, {Vasei}, {Alavi},
  {Giavalisco}, {Siana}, {Grazian}, {Hasinger}, {Suh}, {Cappelluti}, {Vito},
  {Amorin}, {Balestra}, {Brusa}, {Calura}, {Castellano}, {Comastri}, {Fontana},
  {Gilli}, {Mignoli}, {Pentericci}, {Vignali}, \& {Zamorani}}]{Vanzella2016}
{Vanzella}, E., {de Barros}, S., {Vasei}, K., {et~al.} 2016, \apj, 825, 41,
  \dodoi{10.3847/0004-637X/825/1/41}

\bibitem[{{Vanzella} {et~al.}(2018){Vanzella}, {Nonino}, {Cupani},
  {Castellano}, {Sani}, {Mignoli}, {Calura}, {Meneghetti}, {Gilli}, {Comastri},
  {Mercurio}, {Caminha}, {Caputi}, {Rosati}, {Grillo}, {Cristiani}, {Balestra},
  {Fontana}, \& {Giavalisco}}]{Vanzella2018}
{Vanzella}, E., {Nonino}, M., {Cupani}, G., {et~al.} 2018, \mnras, 476, L15,
  \dodoi{10.1093/mnrasl/sly023}

\bibitem[{{Vanzella} {et~al.}(2022){Vanzella}, {Castellano}, {Bergamini},
  {Meneghetti}, {Zanella}, {Calura}, {Caminha}, {Rosati}, {Cupani},
  {Me{\v{s}}tri{\'c}}, {Brammer}, {Tozzi}, {Mercurio}, {Grillo}, {Sani},
  {Cristiani}, {Nonino}, {Merlin}, \& {Pignataro}}]{Vanzella2022}
{Vanzella}, E., {Castellano}, M., {Bergamini}, P., {et~al.} 2022, \aap, 659,
  A2, \dodoi{10.1051/0004-6361/202141590}

\bibitem[{{Verhamme} {et~al.}(2017){Verhamme}, {Orlitov{\'a}}, {Schaerer},
  {Izotov}, {Worseck}, {Thuan}, \& {Guseva}}]{Verhamme2017}
{Verhamme}, A., {Orlitov{\'a}}, I., {Schaerer}, D., {et~al.} 2017, \aap, 597,
  A13, \dodoi{10.1051/0004-6361/201629264}

\bibitem[{{Wang} {et~al.}(2021){Wang}, {Heckman}, {Amor{\'\i}n}, {Borthakur},
  {Chisholm}, {Ferguson}, {Flury}, {Giavalisco}, {Grazian}, {Hayes}, {Henry},
  {Jaskot}, {Ji}, {Makan}, {McCandliss}, {Oey}, {{\"O}stlin}, {Saldana-Lopez},
  {Schaerer}, {Thuan}, {Worseck}, \& {Xu}}]{Wang2021}
{Wang}, B., {Heckman}, T.~M., {Amor{\'\i}n}, R., {et~al.} 2021, \apj, 916, 3,
  \dodoi{10.3847/1538-4357/ac0434}

\bibitem[{{Wyithe} {et~al.}(2011){Wyithe}, {Mould}, \& {Loeb}}]{Wyithe2011}
{Wyithe}, J. S.~B., {Mould}, J., \& {Loeb}, A. 2011, \apj, 743, 173,
  \dodoi{10.1088/0004-637X/743/2/173}

\bibitem[{{Xu} {et~al.}(2016){Xu}, {Wise}, {Norman}, {Ahn}, \&
  {O'Shea}}]{Xu2016}
{Xu}, H., {Wise}, J.~H., {Norman}, M.~L., {Ahn}, K., \& {O'Shea}, B.~W. 2016,
  \apj, 833, 84, \dodoi{10.3847/1538-4357/833/1/84}

\bibitem[{{Yoshiura} {et~al.}(2017){Yoshiura}, {Hasegawa}, {Ichiki}, {Tashiro},
  {Shimabukuro}, \& {Takahashi}}]{Yoshiura2017}
{Yoshiura}, S., {Hasegawa}, K., {Ichiki}, K., {et~al.} 2017, \mnras, 471, 3713,
  \dodoi{10.1093/mnras/stx1754}

\end{thebibliography}
\bibliographystyle{aasjournal}



\end{document}